\documentclass[journal]{IEEEtran} 

\usepackage{fb,afterpage} 
\labelfigure{./}
\graphicspath{{./}}

\usepackage{tikz}
\usetikzlibrary{arrows}
\usepackage[linesnumbered,lined, algoruled]{algorithm2e}

\def\wrt{\hbox{w.r.t. }}
\def\st{\hbox{s.t. }}

\def\pst{\phantom{\st}}
\def\snr{\Omega}
\def\Tom#1{{#1}}

\begin{document}

\title{Coding in the Finite-Blocklength Regime: \\Bounds based on Laplace Integrals \\and their Asymptotic Approximations}

\author{%
Tomaso~Erseghe
\thanks{T. Erseghe is with Dipartimento di Ingegneria dell'Informazione, Universit\`a di Padova, Via G. Gradenigo 6/B, 35131 Padova, Italy. Contact info:  tel +39 049 827 7656, fax +39 049 827 7699, mail erseghe@dei.unipd.it.
} %
\thanks{This work has been submitted to the IEEE for possible publication. Copyright may be transferred without notice,
after which this version may no longer be accessible.}
} 

\maketitle


\acrodef{AWGN}{additive white Gaussian noise} 
\acrodef{BSC}{binary symmetric channel} 
\acrodef{CDF}{cumulative distribution function} 
\acrodef{CCDF}{complementary cumulative distribution function} 
\acrodef{DMC}{discrete memoryless channel} 
\acrodef{FA}{false alarm} 
\acrodef{ISP}{improved sphere-packing} 
\acrodef{i.i.d.}{independent identically distributed} 
\acrodef{LDPC}{low density parity check} 
\acrodef{MD}{missed detection} 
\acrodef{ML}{maximum likelihood} 
\acrodef{LDPC}{low density parity check} 
\acrodef{PAM}{pulse amplitude modulation} 
\acrodef{PPV}{Polyanskyi-Poor-Verd\`u}
\acrodef{PDF}{probability density function} 
\acrodef{RCU}{random coding union} 
\acrodef{SNR}{signal to noise ratio}

\begin{abstract}
In this paper we provide new compact integral expressions and associated simple asymptotic approximations for converse and achievability bounds in the finite blocklength regime. The chosen converse and random coding union bounds were taken from the recent work of Polyanskyi-Poor-Verd\`u, and are investigated under parallel \acs{AWGN} channels, the \acs{AWGN} channels, the BI-\acs{AWGN} channel, and the \acs{BSC}. The technique we use, which is a generalization of some recent results available from  the literature, is to map the probabilities of interest into a Laplace integral, and then solve (or approximate) the integral by use of a steepest descent technique. The proposed results are particularly useful for short packet lengths, where the normal approximation may provide unreliable results.
\end{abstract}


\markboth{SUBMITTED TO IEEE TRANS. ON INFORMATION THEORY}{SUBMITTED TO IEEE TRANS. ON INFORMATION THEORY}

\section{Introduction}

Coding bounds in the finite blocklength regime have recently become quite popular for their ability to capture a compact (and meaningful) description of the physical layer to be used, e.g., for upper layers optimization. These bounds date back to the work of Shannon, Gallager, and Berlekamp \cite{SHANNON1967}, and have \Tom{received new interest} now that powerful coding and decoding techniques that reach the limits of reliable communications are commonly used.

Among the many results available in the \Tom{recent} literature, a widely used practice is to identify the limits of communication for a coding block of size $n$ via the so called \emph{normal approximation}
$$
R\sub{NA} = C - \sqrt{\frac Vn}\,\log_2(e)\,Q^{-1}(P_e) + \frac{\log_2(n)}{2n}\;,
\e{UY2}
$$
where \Tom{$R$ is the rate,} $C$ is the \emph{channel capacity}, $V$ is the \emph{channel dispersion} coefficient, \Tom{and $P_e$ is the average error probability}. The normal approximation \e{UY2} has been proved to be a valid $O(1/n)$ asymptotic approximation for both \emph{achievability} and \emph{converse} bounds \Tom{\cite{altug2010moderate,altug2014moderate, polver10moderate,tan2015it}}, where an achievability bound is intended as a performance that can be achieved by a suitable encoding/decoding couple, while a converse bound is intended as a performance that outperforms any choice of the encoding/decoding couple. Hence \e{UY2} is a good estimate of the limits of information in the finite blocklength regime. Incidentally, Shannon already noticed in 1959 that the normal approximation, but without the $\log_2(n)/2n$ term, applies to the \ac{AWGN} channel case \cite[\S X]{Shannon59}\Tom{, although he established the inverse equation expressing an error probability bound as a function of the rate}. But bound \e{UY2} has also been investigated in \cite{polyanskiy2009dispersion, polyanskiy2010channel,Polyanskiy10,tan2015it} for \ac{AWGN} and parallel \ac{AWGN} channels, the \ac{BSC} channel, the \ac{DMC} channel, and the erasure channel. Upper layer optimization techniques that use \e{UY2} are available in \cite{kim13globecom,makki14globecom,Cent1510:Energy}.

Although \e{UY2} provides a compact and simple description of the limits of information for sufficiently large blocklength $n$, a $O(n^{-1})$ approximation might be unreliable for small $n$, which today corresponds to the rapidly growing scenarios of low-latency machine-to-machine communications. Therefore exact bounds are still of interest for clearly assessing the region of applicability of \e{UY2}, as well as more refined (but simple) approximations are needed to cover the regions where \e{UY2} fails, \Tom{and to get a better theoretical understanding of the limits of communication. The further request of new methodologies, to approach the calculation of achievability and converse bounds in scenarios that were not previously practicable, fully sets} the focus of the paper. \Tom{The recent work of Moulin \cite{moulin2013log} goes in the same direction, using alternative techniques based upon large deviation analysis in order to provide fourth order refinements to \e{UY2}.}

In this paper we investigate some achievability and converse bounds that were recently proposed by Polyansky, Poor, and Verd\'u in \cite{Polyanskiy10}, for which we are able to provide simple integral expressions, for numerical evaluation purposes, and reliable asymptotic approximations that outperform \e{UY2} for small $n$. The significant bounds we investigate are the \ac{PPV} \Tom{meta-}converse bound \cite[\Tom{Theorem~28}]{Polyanskiy10}, and the \ac{RCU} achievability bound \cite[Theorem~16]{Polyanskiy10}. Unlike the approach used by Shannon \cite{Shannon59}, the chosen bounds are built on statistical properties (as opposed to the geometric construction associated with the sphere/cone packing problems), and can therefore be applied to any channel model. They are also known to be consistent with the normal approximation limit, and therefore are expected to be very \Tom{tight}. The results we are proposing are a generalization of the findings available for the \ac{AWGN} channel in \cite{Erseghe15}. \Tom{More specifically, while the derivation in \cite{Erseghe15} relied on a asymptotic uniform series expansion available from the work of  Temme \cite{Temme93}, in the present work we apply the (general) method used by Temme to derive new asymptotic expansions for a number of cases of interest not previously discussed in the literature, which, in turn, will provide new expressions for converse and achievability bounds associated with the finite-blocklength regime. In its essence, t}he leading idea is to give a Laplace integral expression to the probabilities of interest, which can then be solved, or approximated, by using the \emph{steepest descent} method \cite[\S7]{Bleistein75}. \Tom{Some of the tools we are exploiting are in very close relation to those used in the work of Martinez, i Fabregas, et al \cite{martinez2011saddlepoint,martinez2014complex} as well as in the work of Tan and Tomamichel \cite{tan2015it} for evaluating the \ac{RCU} \Tom{achievability} bound, thus certifying the usefulness of the Laplace transform approach in this kind of problems.}

The specific contribution given in this paper covers parallel \ac{AWGN} channels, the standard \ac{AWGN} channel, the so called BI-\ac{AWGN} channel (i.e., binary transmission in an \ac{AWGN} context), and the \ac{BSC}. In parallel \ac{AWGN} channels, 
for which the bound available from the literature is the normal approximation \cite{polyanskiy2010channel, park2012new, tan2015it}, we are able to provide an integral expression and a $O(n^{-3})$ asymptotic approximation to the \ac{PPV} \Tom{meta-}converse bound. For the \ac{AWGN} sub-case, a $O(n^{-2})$ approximation is identified for the \ac{RCU} bound. \Tom{The result improves the ones available in \cite{martinez2011saddlepoint, martinez2014complex} since, although the  Laplace transform approach we are using is similar, we avoid relaxing the \ac{RCU} bound through Markov's inequality and the Chernoff bound. In such a way we are also able to identify an expression which is consistent with (i.e., which simplifies into) the normal approximation \e{UY2}.} For the BI-\ac{AWGN} channel we are able to provide an integral expression and a $O(n^{-3})$ asymptotic approximation to the \ac{PPV} \Tom{meta-}converse bound which significantly outperform the state-of-the-art bounds available from Shannon \cite{SHANNON1967}, Valembois and Fossorier \cite{Valembois:it04}, and Wiechman and Sason \cite{wiechman:it08}. Although the technique used in this paper is not able to deal with the \ac{RCU} achievability bound, use of the weaker $\kappa\beta$ achievability bound of \cite[Theorem~25]{Polyanskiy10} clearly reveals the $1\,$dB gap existing between what we could achieve and the performance of a standard \emph{message passing} decoder, the gap being valid over a very large blocklength range (i.e., for short as well as for long codes). \Tom{It is anyway worth mentioning that for very short packet sizes, one can construct codes (and decoders) that approach the achievability bound, and in some regions (high error rate) even beat it \cite{liva2013short}.} For the \ac{BSC} we are finally providing reliable approximations to both the \ac{PPV} \Tom{meta-}converse and the \ac{RCU} achievability bounds, thus improving over the results available in \cite{Polyanskiy10} and based upon a series expression. Overall, all the above results show that the \ac{RCU} and the \ac{PPV} \Tom{meta-converse} bounds are both consistent with the normal approximation (hence asymptotically optimum), that the \ac{PPV} \Tom{meta-converse} bound is in general neater to calculate and that it can be therefore taken as (a very good approximation to) \emph{the} performance limit, at least for $n\ge200$ where upper and lower limits are sufficiently close. With \Tom{proper} modifications, which are not discussed in this paper \Tom{and which are left for future investigation}, our techniques can also be applied to other channels of practical interest, namely the \ac{DMC} channel, or the \ac{AWGN} channel under a specific modulation/constellation choice.

The rest of the paper is organized as follows. The \ac{PPV} \Tom{meta-converse}, \ac{RCU}, and $\kappa\beta$ bounds used in this paper are defined in \sect{BO}\Tom{, and a brief description of the Laplace method used is available in \sect{ME}. The bounds} are evaluated in the four scenarios of interest in successive sections: \sect{PA} deals with the parallel \ac{AWGN} case, \sect{GA} deals with the simple \ac{AWGN} case, \sect{SO} deals with the BI-\ac{AWGN} channel, and \sect{HA} deals with the \ac{BSC}. To keep the flow of discussion, all theorems proofs (or proof sketches) are made available in the Appendix. \Tom{Fort the sake of readability, the most significant results are summarized, at the end of each section, in a compact procedural form. }

\Section[BO]{The Bounds}

In the following we will consider two fundamental results, namely, the \ac{PPV} \Tom{meta-converse} bound of \cite[\Tom{Theorem~28}]{Polyanskiy10} which is a converse bound setting a limit to the best obtainable performance, and the \ac{RCU} bound \cite[Theorem~16]{Polyanskiy10} which is an achievability bound setting a performance that can be obtained by a suitable encoding/decoding technique. \Tom{The considered performance measure is the \emph{average} error probability $P_e$, but we warn the reader that some of the following results (i.e., the \ac{PPV} \Tom{meta-converse} and the related $\kappa\beta$ bounds) are also applicable in the \emph{maximum} error probability sense.}

By using the notation of \cite{Erseghe15}, the converse bound of interest to our investigation can be defined as follows.

\begin{theorem}[\ac{PPV} \Tom{meta-converse} bound]\label{theo1}
Assume a codebook $\C C$ with $M$ codewords of length $n$, with associated rate $R=(\log_2M)/n$, and further let the codewords belong to set $\C K$. Assume that codewords are transmitted over a channel described by the transition \ac{PDF} $p_{y|x}(\B b|\B a)$, and denote with $P_e$ the average error probability (in decoding) when symbols are equally likely. Choose a (arbitrary) output \ac{PDF} $p_y(\B b)$, and define the log-likelihood function
$$
\Lambda(\B a, \B b) = \frac1n\ln \frac{p_{y|x}(\B b|\B a)}{p_y(\B b)}\;.
\e{KK10}
$$
Denote the Neyman-Pearson \ac{MD} and \ac{FA} probabilities respectively as
$$
\eqalign{
P\sub{FA}(\B a,\lambda) & = \P{\Lambda(\B a,\B y)\ge\lambda},\; \Tom{\hbox{with } \B y\sim p_y}\cr
P\sub{MD}(\B a,\lambda) & = \P{\Lambda(\B a,\B y)<\lambda},\; \Tom{\hbox{with }\B y\sim p_{y|x}, \B x=\B a}\;.
}
\e{KK12}
$$
If probabilities \e{KK12} are independent of $\B a$ for $\B a\in\C K$, then for a fixed error probability $P_e$ the rate $R$ is upper bounded by
$$
R  \le \overline{R} = -\fract1n\log_2(P\sub{FA}(\lambda))\;,
\e{KK14}
$$
where $\lambda$ is set by the constraint $P\sub{MD}(\lambda) = P_e$.\hfill~$\Box$
\end{theorem}
\begin{IEEEproof}
See \cite[\Tom{Theorem~28}]{Polyanskiy10} and \cite{Erseghe15}.
\end{IEEEproof}

Reading the result the other way round, for a fixed rate $R$ the error probability $P_e$ is lower bounded by
$$
P_e  \ge \underline{P}_e = P\sub{MD}(\lambda)\;,
\e{KK14b}
$$
where $\lambda$ is set by the constraint $P\sub{FA}(\lambda) = 2^{-nR}$. We also note that, although the theorem is defined for the average error probability, its bounds are applicable also in the maximum error probability case. This is intuitively explained by the fact that \Tom{the maximum error probability is by definition greater than the average error probability, hence an upper bound established for a given average $P_e$ corresponds to a bound that holds for a maximum error probability which is greater than $P_e$ (and a similar rationale applies to \e{KK14b}). We finally observe that} the natural choice for $p_y(\B b)$ is the capacity achieving expression which, as we will see, fully satisfies the theorem request on Neyman-Pearson probabilities (i.e., they are independent of $\B a$) in all the case of practical interest \Tom{to this paper}, and it further sets the bound to capacity for large values of $n$. \Tom{In general, however, other densities $p_y(\B b)$ might be used for obtaining a meaningful result.}

For the achievability bound we have the following result.

\begin{theorem}[\ac{RCU} \Tom{achievability} bound]\label{theo21} By using the notation of Theorem~\ref{theo1}, the best achievable average error probability is upper bounded by expression
$$
P_e \le \overline{P}_e =  \E{ \min(1,2^{nR}g(\B x,\B y))},\Tom{\hbox{with } \B y\sim p_{y|x}, \B x\sim\C U_{\C K}},
\e{DS2}
$$
where
$$
g(\B x,\B y) = \P{\Lambda(\B z, \B y)\ge\Lambda(\B x, \B y)},\;\Tom{\hbox{with } \B z\sim \C U_{\C K}}\;,
\e{DS44}
$$
and where $\C U_{\C K}$ denotes a uniform distribution over set $\C K$.~\hfill$\Box$
\end{theorem}
\begin{IEEEproof}
See \cite[Theorem~16]{Polyanskiy10}.
\end{IEEEproof}

\Tom{Although, in general, any distribution can be chosen for both $\B z$ and $\B x$, and not necessarily the uniform distribution $\C U_{\C K}$, the results of this paper all rely on a uniform distribution bound.} Observe also that, similarly to the \ac{PPV} \Tom{meta-converse} case, from \e{DS2} we can obtain, for any given error probability $P_e$, a lower bound on the best achievable rate, $R\ge\underline{R}$. Also, the natural choice for $p_y(\B b)$ will again be the capacity achieving expression, which, in the \Tom{scenarios of interest to this paper}, will guarantee an independence of the function in \e{DS2} of the specific choice of $\B x$. 

Since the derivation of the \ac{RCU} bound may be challenging in many occasions, we also introduce the (weaker) $\kappa\beta$ bound proposed by \cite{Polyanskiy10}, which relies on the \ac{MD} and \ac{FA} probabilities.

\begin{theorem}[$\kappa\beta$ bound]\label{theo221}
By using the notation and the assumptions of Theorem~\ref{theo1}, the best achievable rate is lower bounded by expression
$$
R\ge \underline{R} = \max_{\tau\in(0,P_e)} \fract1n\log_2(\kappa(\tau)) - \fract1n\log_2( P\sub{FA}(\lambda))
\e{KB2}
$$
where $\lambda$ is set by the constraint $P\sub{MD}(\lambda) = P_e-\tau$, and where
$$
\eqalign{
\kappa(\tau) = & \min_{\C R} \int_{\C R} p_y(\B b)\;d\B b\cr
& \st \int_{\C R^c} p_{y|x}(\B b|\B a)\;d\B b \le 1-\tau \;,\forall \B a\in\C K
}
\e{KB4}
$$
for some choice of region $\C R\subset \M R^n$, and with $\C R^c$ the complement region.~\hfill$\Box$
\end{theorem}
\begin{IEEEproof}
See \cite[Theorem~25]{Polyanskiy10}.
\end{IEEEproof}

 \Tom{Although the $\kappa\beta$ bound is naturally a bound on the \emph{maximum} error probability, it is also applicable in an \emph{average} error probability sense. As for the \ac{PPV} \Tom{meta-converse} bound case, this is due to the fact that the maximum error probability is by definition greater than the average error probability, hence a lower bound established for a given maximum error probability $P_{e,\rm max}$ corresponds to a bound that holds for an average error probability that satisfies $P_e\le P_{e,\rm max}$. This is also the reason why, in the average error probability sense, the $\kappa\beta$ bound is weaker than the \ac{RCU} bound.}
 
We finally note that the \ac{PPV} \Tom{meta-converse} and the $\kappa\beta$ bounds of Theorem~\ref{theo1} and Theorem~\ref{theo221} are correct under the assumption that $\B y$ is a continuous distribution, and require some modifications in the case of discrete variables. Modifications to the reference results will be discussed where needed (i.e., in the \ac{BSC} case of \sect{HA}).

\Section[ME]{The Method}

\Tom{Before delving into the derivation of the bounds, we briefly review the general method which will be used throughout the paper, with appropriate modifications depending on the specific problem considered from time to time. The method is taken from \cite{Temme93}, and suitably adapted to all the considered scenarios. The leading idea is that a probability of the form
$$
P = \P{\sum_{i=1}^n u_i \le n \lambda}
\e{ME2}
$$
where $u_i$'s are \ac{i.i.d.} continuous random variables with \ac{PDF} $f_{u}(a)$, can be efficiently written (and numerically evaluated) by using standard Laplace transform properties (e.g., see \cite{oppenheim2014signals}). As a matter of fact, summing \ac{i.i.d.} random variables corresponds to convolving their \acp{PDF}, that is multiplying the corresponding Laplace transforms. Hence,  \e{ME2} can be written in the equivalent (inverse Laplace transform) form
$$
\eqalign{
P & = \frac1{i2\pi} \int_{\C L} \frac{(F_u(s))^n e^{n\lambda s}} s ds\cr
 & =  \frac1{i2\pi} \int_{\C L} \frac{e^{\frac n2\alpha(s)}} s ds \;,\quad \alpha(s) = 2\ln(F_u(s))+2\lambda s
}
\e{ME4}
$$
where $F_u(s)$ is the bilateral Laplace transform of $f_{u}(a)$, namely
$$
F_u(s) = \int_{-\infty}^{+\infty} f_{u}(a) e^{-s a}\,da\;,\quad s\in\C R
\e{ME6}
$$
with $\C R\subset\M C$ the \emph{region of convergence} of the integral.}

\Tom{The standard integration path $\C L$ in \e{ME4} is any line of the form $\C L=\{s|\Re(s)=\gamma\in \C R\}$, but can be modified into a more convenient integration path. A good choice is to select a path where the Laplace transform does not oscillate, in such a way to  simplify (and strengthen) the numerical evaluation of the integral. This corresponds to using a path $\C D$ which transits through one \emph{saddle point} $s_\alpha$ of the function $\alpha(s)$ (see \e{ME4}), i.e. through one zero of $\alpha'(s)$,  and such that the imaginary part of $\alpha(s)$ is constant in $\C D$. Since $\alpha(s)$ is analytic by construction, hence the Cauchy--Riemann equations apply, then the resulting path is ensured to be a \emph{steepest descent} integration path, i.e., one where $\alpha(s)$ decreases (e.g., see \cite[\S7]{Bleistein75}). This results in
$$
P = \frac1{i2\pi} \int_{\C D} \frac{e^{\frac n2\alpha(s)}} s ds\;,
\e{ME8}
$$
where $\C D$ has the form $\C D =\{s| \Im(s)=\Im(s_\alpha), \alpha'(s_\alpha)=0\}$ or, more generally, is a connection of a number of descent paths. We warn the reader that an appropriate use of the Theorem of Residues must be considered in \e{ME8} in case the selected path $\C D$ crosses some of the poles of $\alpha(s)$.}

\Tom{Although \e{ME8} is good enough for numerical integration purposes (mainly because of the non-oscillating nature of $\alpha(s)$ on $\C D$), some further properties can be exploited to obtain an asymptotic expansion as in \cite{Temme93}. This can be done by relying on the descending nature of $\alpha(s)$ on $\C D$, in such a way to parametrize the path $\C D$ in the form $\C D= \{p_\alpha(u), u\in[a,b]\}$ where the complex map $p_\alpha(u)\in \C D$ is such that
$$
\alpha(p_\alpha(u)) = \alpha(s_\alpha) - u^2\;.
\e{ME10}
$$
With the newly introduced notation we can write \e{ME8} in the form
$$
P = \frac1{i2\pi} e^{\frac n2\alpha(s_\alpha)} \int_{a}^b \frac{p'_\alpha(u)} {p_\alpha(u)} e^{ - \frac n2u^2}  du\;,
\e{ME12}
$$
where the path derivative is (from \e{ME10})
$$
p'_\alpha(u) = \frac{2u}{-\alpha'(p_\alpha(u))}\;.
\e{ME14}
$$
Appropriate $O(n^{-k})$ asymptotic expansions can then be obtained by approximating the function $p'_\alpha(u)/p_\alpha(u)$ with its  truncated Taylor expansion at $u=0$, and by recalling that the integral $\int_a^b u^k e^{ - \frac n2u^2}  du$ is solvable in the closed form by using the (lower) incomplete Gamma function. The coefficients of the Taylor expansion will be evaluated, using standard methods, either in the closed form or in numerical form, depending on the availability of a closed form expression for $\alpha(s)$.}

\Section[PA]{Parallel Gaussian Channels}

\subsection{Scenario and notation}

In the parallel Gaussian channel scenario the codeword $\B x\in\C C$, of length $n$, is assumed to be partitioned in $K$ \Tom{parallel channels, that is $K$ blocks, $\B x_k$, $k=1,\ldots,K$, of equal length $n/K$. Hence, in the following $n$ is assumed to be an integer multiple of $K$. In the chosen scenario,} codewords are assumed to have constant-power in each block, that is $\B x\in \C K $ with set $\C K$ defined as
$$
\C K = \Big\{ \B x = [\B x_1,\ldots,\B x_K] \Big| \|\B x_k\|^2 = \Tom{\fract nK P_k}  \Big\}
\e{KK1}
$$
where $P_k$ is the average power associated with the $k$th block. \Tom{Note that we are therefore considering a situation where the bounds are evaluated for a specific power vector $\{P_1,\ldots,P_K\}$, which simplifies equations, and which will not prevent us to lately apply power availability constraints of the form
$$
\sum_{k=1}^K P_k = P\sub{tot}\;,
\e{KK1bis}
$$
by simple use of standard constrained optimization techniques (e.g., see the \emph{water-filling} application in Sect.~\ref{wfil2}). Our choice is dictated by the fact that a closed-form identification of the optimum power assignment under constraint \e{KK1bis} is not available in general, except made for the normal approximation discussed in \cite[\S 4]{polyanskiy2010channel} where the standard \emph{water-filling} solution can be used.}

On the transmission side each block sees a different \ac{AWGN} channel. The \ac{AWGN} channel experienced by the $k$th block has the form $\B y_k=\B x_k+\B w_k$, with $\B w_k\sim\C N(\B 0,\B I\sigma_k^2)$ a zero-mean Gaussian noise vector with independent entries. The corresponding \ac{SNR} is $\snr_k=P_k/\sigma_k^2$. Channel transition \acp{PDF} are therefore of the form
$$
\eqalign{
p_{y|x}(\B b|\B a) & = \prod_{k=1}^K \frac1{(2\pi \sigma_k^2)^{\frac12  \Tom{\frac nK}}} \exp\left(-\frac{\|\B b_k-\B a_k\|^2}{2\sigma_k^2}\right)\;.
}
\e{KK4}
$$
Receive \acp{PDF} $p_{y}(\B b)$ are also of interest. By assuming capacity achieving Gaussian inputs $\B x_k\in\C N(\B 0,\B IP_k)$, we have
$$
\eqalign{
p_{y}(\B b) & = \prod_{k=1}^K \frac1{(2\pi \sigma_k^2 (1+\snr_k))^{\frac12  \Tom{\frac nK}}} \exp\left(-\frac{\|\B b_k\|^2}{2\sigma_k^2(1+\snr_k)}\right)\;,
}
\e{KK6}
$$
\Tom{which denotes an auxiliary output distribution that depends weakly on the input codeword through the presence of the \ac{SNR} contribution $\snr_k$. In this context, being the powers $\{P_k\}$ fixed, capacity is simply expressed by} 
$$
C = \Tom{\frac1K} \sum_{k=1}^K \fract12\log_2(1+\snr_k)\;.
\e{KK64}
$$

\Tom{As a concluding comment, we observe that in \e{KK1} we are considering a scenario where the power is fixed to a given level. Generalizations to maximum power and average power settings can be obtained by using the results of  \cite[\S 4]{polyanskiy2010channel}.}

\subsection{\ac{PPV} \Tom{meta-converse} converse bound}

In order to be able to calculate the \ac{PPV} \Tom{meta-converse} bound of Theorem~\ref{theo1}, and to derive suitable asymptotic approximations, we can mimick the asymptotic expansion approach of Temme \cite{Temme93} and the results of  \cite{Erseghe15}, both valid in a standard Gaussian (non parallel) channel case. The leading idea is to develop a simple integral form in the Laplace domain, which can then be used to define an asymptotic expansion by means of the method of steepest descent (e.g., see \cite[\S7]{Bleistein75}), as we explain in the following.

We start our derivation by showing that, in a parallel \ac{AWGN} channels scenario, the \ac{FA} and \ac{MD} probabilities \e{KK14} evaluated under \e{KK4}-\e{KK6} satisfy the assumptions of Theorem~\ref{theo1}. Specifically they are independent of the value of $\B a$ provided that $\B a\in\C K$, and take a simple form which depends on non-central chi-squared distributions, as stated by the following result.

\begin{theorem}\label{theo2}
In a parallel \ac{AWGN} channels scenario the \ac{FA} and \ac{MD} probabilities \e{KK12} are independent of the specific value of $\B a$, and are given by
$$
\eqalign{
P\sub{FA}(\lambda) & = \P{\sum_{k=1}^K  u_k\,\snr_k\le  n\lambda}\cr
P\sub{MD}(\lambda) & = \P{\sum_{k=1}^K \frac{ v_k\,\snr_k}{1+\snr_k}\!> n\lambda}\cr
}
\e{KK15}
$$
where
$$
\Tom{\eqalign{
u_k &\sim\chi\Big( \Tom{\frac nK},\Tom{\frac nK} \frac{1+\snr_k}{\snr_k}\Big)\;,\quad
v_k  \sim\chi\Big(\Tom{\frac nK},\Tom{\frac nK}\frac{1}{\snr_k}\Big)
}}
\e{KK15seconda}
$$
and where $\chi(n,s)$ denotes a non-central chi-squared random variable of order $n$ and parameter $s$.\hfill~$\Box$
\end{theorem}
\begin{IEEEproof}
See the Appendix.
\end{IEEEproof}

Starting from this result, a compact integral expression for the probabilities of interest can be derived by use of standard Laplace transform properties.

\begin{theorem}\label{theo3}
In a parallel \ac{AWGN} channels scenario the \ac{FA} and \ac{MD} probabilities \e{KK12} can be expressed in the form
$$
\eqalign{
P\sub{FA}(\lambda)  &= \frac{1}{i2\pi} \int_{\gamma-i\infty}^{\gamma+i\infty} \frac{e^{\frac n2\alpha(s)}}{s}  \,ds\cr
P\sub{MD}(\lambda)  &= \frac{1}{i2\pi} \int_{\mu-i\infty}^{\mu+i\infty} \frac{e^{\frac n2\beta(s)}}{-s}  \,ds\cr
}
\e{KK30}
$$
where 
$$
\eqalign{
\alpha(s) 
& =2\lambda s -  \Tom{\frac1K}\sum_{k=1}^K  \left(\frac{2s(1+\snr_k)}{1+2 s \snr_k}+\ln(1+2 s \snr_k)\right) \cr
\beta(s) 
& = \alpha(s+\fract12) +  2\ln(2) \cdot C +1-\lambda 
}
\e{KK32a}
$$
and where $\gamma>0$ and $0>2\mu>-1-1/\max_k\snr_k$.\hfill~$\Box$
\end{theorem}
\begin{IEEEproof}
See the Appendix.
\end{IEEEproof}

Observe that, for $K=1$, the result of Theorem~\ref{theo3} is equivalent to the findings of \cite{Temme93} (and \cite{Erseghe15}), although with a different notation. This is also true for the results presented later on in the text and referring to the case $K=1$.

\subsection{Applying the method of steepest descent}

Efficient solution of the integrals in \e{KK30} can be achieved by appropriately changing the integration path in such a way that it corresponds to a steepest descent path \cite[\S7]{Bleistein75}. This requires identifying suitable saddle points, that is, points $s_\alpha$ and $s_\beta$ such that $\alpha'(s_\alpha)=0$ and $\beta'(s_\beta)=0$. The function symmetry further ensures that saddle points are real valued. For the sake of easier tractability, we also require saddle points to belong to the Laplace regions of convergence which generated the exponential contributions in \e{KK30}, that is to sets (see also the proof of Theorem~\ref{theo3} for details)
$$
\eqalign{
\C S_\alpha & = \Big\{ s\Big| \Re(s)>-\fract1{2\snr_k}, k=1,\ldots,K  \Big\}\cr
\C S_\beta  & = \C S_\alpha-\fract12\;,\cr
}
\e{KK43}
$$
in such a way that poles of $e^{\frac n2\alpha(s)}$ and $e^{\frac n2\beta(s)}$ are not crossed by the new integration path. This is a viable option, and because of convexity of $\alpha(s)$ in $\C S_\alpha\cap\M R$ we also have
$$
s_\alpha = \argmin_{s\in\C S_\alpha\cap\M R}\alpha(s)\;,
\e{KK43a}
$$
which can be efficiently solved via standard convex optimization methods. Alternatively one can solve
$$
 \Tom{\frac1K} \sum_{k=1}^K  \frac{1+2\snr_k+2s_\alpha\snr_k^2}{(1+2 s_\alpha \snr_k)^2} = \lambda\;.
\e{KK43aB}
$$
From the second of \e{KK32a} it is then evident that the saddle point $s_\beta$ simply satisfies 
$$
s_\beta=s_\alpha-\fract12\;.
\e{KK43a2}
$$ 
We also note that a closed form result for \e{KK43a} is available only in the \ac{AWGN} case, $K=1$, providing
$$
s_\alpha=  -\frac1{2\snr} + \frac{1}{4\lambda} \Big(1+\sqrt{1+4\lambda\fract{(1+\snr)}{\snr^2}}\Big)\;.
\e{KK60}
$$

The steepest descent path $\C L_\alpha$ is then identified by the equivalence $\Im(\alpha(s))=\Im(\alpha(s_\alpha))=0$, with the additional request to include the saddle point $s_\alpha$ when crossing the real axis, which ensures uniqueness of the path. The path $\C L_\beta$ is instead simply given by
$$
\C L_\beta = \C L_\alpha - \fract12\;,
\e{DS6b}
$$
which is again a straightforward consequence of the definition of $\beta$ in \e{KK32a}. We also observe that, because of the symmetry of function $\alpha(s)$, steepest descent paths are symmetric with respect to the real axis, which, incidentally, correctly ensures the integrals \e{KK30} be imaginary valued, and probabilities real valued. Hence, the steepest descent path  $\C L_\alpha$ can be parametrized by means of a variable $u\in\M R$ such that $\alpha(s) = \alpha(s_\alpha) - u^2$ and ${\rm sign}(u)={\rm sign}(\Im(s))$, that is
$$
u = f(s) = {\rm sign}(\Im(s)) \;\sqrt{\alpha(s_\alpha)-\alpha(s)}\;,\quad s\in\C L_\alpha\;.
\e{DS14}
$$
Operatively, this defines the path in the form 
$$
\C L_\alpha=\{p_\alpha(u), \;u\in\M R\}\;, 
\e{YO2}
$$
with map $u\rightarrow p_\alpha(u)=s$ identified for $u\ge0$ by constraints  
$$
\eqalign{
\Im(s) & \ge 0 \cr
\alpha(s) & = \alpha(s_\alpha) - u^2\cr
\Im(\alpha(s)) & = 0 \cr
}
\e{YO4}
$$
while for $u<0$ it simply is $p_\alpha(u)=p_\alpha^*(-u)$. It also naturally is $p_\alpha(0)=s_\alpha$. Alternatively, the constraint $\Im(\alpha(s))=0$ in \e{YO4} can be expressed in the (equivalent) explicit form
$$
\lambda  =  \Tom{\frac1K}\sum_{k=1}^K \left( \rule{0mm}{6mm}\frac{1\!+\!\snr_k}{|1\!+\!2s\snr_k|^2} 
	+ \frac{1}{\Im(2s)}\cot^{-1}\!\!\left(\frac{\Re(2s)\!+\!\frac1{\snr_k}}{\Im(2s)}\right)\!\! \right)\;,
\e{YO5}
$$
which, with some effort, reveals that $\Re(p_\alpha(u))\le s_\alpha$, and that the edge points are $p_\alpha(\pm\infty)=-\infty\pm i\pi/(2\lambda)$. Unfortunately, a closed form expression is not available for path $p_\alpha(u)$, and we must resort to numerical methods. Nevertheless, this will not prevent us in identifying an asymptotic expansion in the closed form.  

By a change of the integration variable in \e{KK30}, the above ensures validity of the following result.

\begin{theorem}\label{theoN4}
In a parallel \ac{AWGN} channels scenario the \ac{FA} and \ac{MD} probabilities \e{KK12} can be expressed in the form
$$
\eqalign{
P\sub{FA}(\lambda)  &= 1(-s_\alpha) +  \frac1{i2\pi}e^{\frac n2\alpha(s_\alpha)}\int_{-\infty}^{\infty} c_\alpha(u) e^{-\frac n2u^2}  \,du\cr
P\sub{MD}(\lambda)  &= 1(s_\beta) -\frac1{i2\pi}e^{\frac n2\beta(s_\beta)} \int_{-\infty}^{\infty}  c_\beta(u)e^{-\frac n2u^2}  \,du\cr
}
\e{KK40}
$$
where $1(\cdot)$ is the unit step function, where
$$
c_x(u) = \frac{p'_x(u)}{p_x(u)} = \frac{2u}{-\alpha'(p_\alpha(u))\,p_x(u)} = -c^*_x(-u)\;, 
\e{KK40a}
$$
and where $x$ stands for either $\alpha$ or $\beta$.  The path in $\alpha$ is derived, for $u\ge0$, according to \e{YO4}  where $s=p_\alpha(u)$, and for $u<0$ it is extended by symmetry $p_\alpha(u)=p_\alpha^*(-u)$. The path in $\beta$ is $p_\beta(u)=p_\alpha(u)-\frac12$.\hfill~$\Box$
\end{theorem}
\begin{IEEEproof}
See the considerations above.
\end{IEEEproof}

The anti-Hermitian symmetry of $c_x(u)$ stated in \e{KK40a} can be further exploited in order to identify a real valued integral in the form
$$
\eqalign{
P\sub{FA}(\lambda)  & = 1(-s_\alpha) +  \frac1\pi e^{\frac n2\alpha(s_\alpha)}\int_{0}^{\infty} \Im[c_\alpha(u)] e^{-\frac n2u^2}  \,du
\cr
P\sub{MD}(\lambda)  &= 1(s_\beta) -\frac1\pi e^{\frac n2\beta(s_\beta)} \int_{0}^{\infty}  \Im[c_\beta(u)] e^{-\frac n2u^2}  \,du\;.
}
\e{KK41}
$$ 
Incidentally, note that the unit step functions in \e{KK40} and \e{KK41} simply take into account on wether the integration paths are crossing the pole at $0$, in which case the Theorem of Residues was used to correctly identify the result. The expression in \e{KK40a} is instead a straightforward consequence of the derivative of the inverse in \e{DS14}. \Tom{A summary of the overall procedure is available in \fig{FI2}.}

We finally observe that, with a little effort, in the Gaussian case $K=1$ we are able to identify from \e{YO5} the steepest descent path $\C L_\alpha$ in the parametric form
$$
p_\alpha(\phi)  = \frac{e^{j\phi}}{4\lambda\sinc(\phi)} \left(1\!+\!\sqrt{1\!+\!4\lambda\fract{(1+\snr)}{\snr^2}\sinc^2(\phi)}\right)-\frac1{2\snr}\;,
\e{DS4}
$$
where $\sinc(\phi)=\sin(\phi)/\phi$, and $\phi\in(-\pi,\pi)$.

\subsection{Asymptotic expansion}

When the map $p_\alpha(u)$ is known either numerically or in the closed form, then the integral form of Theorem~\ref{theoN4} can be readily used for numerical evaluation since the derivative of $\alpha(s)$ is known from \e{KK32a}. As an alternative, for sufficiently large values of $n$ we can resort to a (tight) asymptotic expansion.

The asymptotic expansion is found by using a Taylor series expansion at $u=0$ for the functions $c_\alpha$ and $c_\beta$, and by exploiting \cite[eq.~2.3.18.2]{Prudnikov_vol1}. As we anticipated, the Taylor series coefficients in \e{KK44} can be evaluated in the closed form even in the absence of a closed form expression for the steepest descent paths, since the derivatives of $p_x(s)$ can be inferred from (inversion of) \e{DS14}. The result can be formulated as follows.

\begin{procedura}\label{theo5}
In a parallel \ac{AWGN} channels scenario the \ac{FA} and \ac{MD} probabilities \e{KK12} allow the asymptotic expansion
$$
\eqalign{
P\sub{FA}(\lambda)  &=  1(-s_\alpha) + \frac{ e^{\frac n2\alpha(s_\alpha)}}{\sqrt{2\pi n}} \sum_{k=0}^\infty    \frac{(2k)!}{k! (2n)^k}  \frac{c_{\alpha,2k}}{i} \cr
P\sub{MD}(\lambda)  &= 1(s_\beta)  + \frac{ e^{\frac n2\beta(s_\beta)}}{\sqrt{2\pi n}} \sum_{k=0}^\infty    \frac{(2k)!}{k! (2n)^k}  \frac{c_{\beta,2k}}{-i} 
}
\e{KK44}
$$
where $c_{\alpha,k}$ and $c_{\beta,k}$ are the Taylor expansion coefficients of the functions $c_\alpha$ and $c_\beta$, respectively, as given in \e{KK40a}. Operatively, for the \ac{FA} probabilities the following iterative procedure can be used to identify the coefficients $c_{\alpha,2k}$:
\begin{enumerate}
\item Evaluate the real valued Taylor series coefficients of the function $\alpha(s)$ in $s=s_\alpha$ \Tom{according to rule:} $a_0=\alpha(s_\alpha)$, $a_1=0$, and 
$$
a_m  =  \Tom{\frac1K}\sum_{k=1}^K   \left(\frac{(1+\snr_k)/\snr_k}{1+2s_\alpha\snr_k}+\frac1m\right)
	\cdot\left(\frac{-2\snr_k}{1+2s_\alpha\snr_k}\right)^m
\e{DG2}
$$
for $m\ge2$. Note that derivatives follow an alternating sign rule, and $a_2\ge0$.
\item Evaluate the imaginary valued Taylor series coefficients  in $s=s_\alpha$ of the function $f(s)=p_\alpha^{-1}(s)$ (defined in \e{DS14}) \Tom{according to rule:} $f_0=0$, $f_1 = -i\sqrt{a_2}$, and 
$$
f_m = -\frac1{2f_1}\left(a_{m+1} + \sum_{\ell=2}^{m-1} f_\ell f_{m+1-\ell}\right)\;.
\e{KK48}
$$
\item Evaluate the Taylor series coefficients  in $u=0$ of the function $p_\alpha(u)$ \Tom{according to rule:} $p_0=s_\alpha$, $p_1=1/f_1$, and
$$
p_m = -\frac1{f_1^m} \sum_{\ell=1}^{m-1} p_\ell P_{\ell,m}
\e{KK50}
$$
with coefficients $P_{\ell,m}$ defined by $P_{m,m}=f_1^m$ and
$$
P_{\ell,m} = \sum_{k=1}^{m-\ell} \frac{(k\ell-m+k+\ell)}{(m-\ell)} \frac{f_{k+1}}{f_1} P_{\ell,m-k}\;.
\e{KK52}
$$
The even coefficients $p_{2m}$ are real valued, and the odd coefficients $p_{2m+1}$ are imaginary valued.
\item Evaluate the Taylor series coefficients in $u=0$ of the function $c_\alpha(u)$ (defined in \e{KK40a}) \Tom{according to rule:}
$$
c_{\alpha,m} = \frac1{ s_\alpha} \left( (m+1) p_{m+1} - \sum_{\ell=1}^m p_\ell\, c_{\alpha,m-\ell} \right)\;.
\e{KK54}
$$
The even coefficients $c_{\alpha,2m}$ are imaginary valued, and the odd coefficients $c_{\alpha,2m+1}$ are real valued.
\end{enumerate}
The coefficients $c_{\beta,2k}$ for the \ac{MD} probabilities can be found similarly, that is:
\begin{enumerate}
\item Evaluate the real valued Taylor series coefficients of the function $\beta(s)$ in $s=s_\beta$ \Tom{according to rule:} $b_0=\beta(s_\beta)$, $b_1=0$, and $b_m=a_m$ as given by \e{DG2} for $m\ge2$.
\item[2-4)] Use the method defined for the \ac{FA} probabilities by replacing $\alpha\rightarrow \beta$ and $a_m\rightarrow b_m$.
 \hfill~$\Box$
\end{enumerate}
\end{procedura}
\begin{IEEEproof}
See the Appendix.
\end{IEEEproof}

\subsection{Reliable approximations}

\Tom{Procedure~\ref{theo5}} is general, in that it provides a method to identify the Taylor coefficients of $c_x(u)$ of any order. However, the fact that the function $c_x(u)$ in \e{KK41} is weighted by $e^{-\frac n2 u^2}$ ensures that only the first coefficients are needed in practice to obtain (for sufficiently large $n$) a very reliable result. In this context, it is of interest investigating both the first-term and second-term approximations. We \Tom{specify} them by assuming that
$$
s_\beta < 0 < s_\alpha
\e{AS2}
$$
holds, which implies the absence of the contribution of the residues $1(-s_\alpha)$ and $1(s_\beta)$. In turn, this corresponds to $P\sub{FA}<\frac12$ and $P\sub{MD}<\frac12$, i.e., to neglecting the cases of very limited interest where $P_e>\frac12$ (too large error probability) and $R<\frac1n$ (single symbol codebook, $M=1$). Hence, by limiting the series to the first terms we obtain the following result.

\begin{corollary}\label{coro02}
In a parallel \ac{AWGN} channels scenario the \ac{FA} and \ac{MD} probabilities \e{KK12} allow the $O(n^{-2})$ asymptotic approximations
$$
{\eqalign{
\frac1n \ln P\sub{FA}^{(1)} &= \fract12 \alpha(s_\alpha) - \frac1{2n}\ln\left(2\pi n a_2 s_\alpha^2\right) \cr
\frac1n\ln P\sub{MD}^{(1)}  & = \fract12 \beta(s_\beta) - \frac1{2n}\ln\left(2\pi n a_2 s_\beta^2\right)\;,
}}
\e{HG3}
$$
and the $O(n^{-3})$ asymptotic approximations
$$
{\eqalign{
\frac1n \ln P\sub{FA}^{(2)} &=\frac1n \ln P\sub{FA}^{(1)} 
	+ \frac1n\ln\Big(1+ \frac1ng(s_\alpha)\Big)  \cr
\frac1n\ln P\sub{MD}^{(2)}  & =  \frac1n\ln P\sub{MD}^{(1)} 
	+ \frac1n\ln\Big(1+ \frac1ng(s_\beta)\Big)  \;,
}}
\e{HG5}
$$
where 
$$
g(s) =  \frac{12a_2a_4-15 a_3^2}{8a_2^3} - \frac{3a_3}{2a_2^2s} - \frac1{a_2s^2}\;,
\e{HG7}
$$
and where the functions $\alpha(s)$ and $\beta(s)$ were defined in \e{KK32a}, $s_\alpha$ was defined in \e{KK43a}, $s_\beta=s_\alpha-\fract12$, and $a_2,a_3,a_4$ were defined in \e{DG2}. The approximations hold provided that \e{AS2} is satisfied. \hfill~$\Box$
\end{corollary}
\begin{IEEEproof}
This is a straightforward application of \Tom{Procedure~\ref{theo5}}.
\end{IEEEproof}

\Tom{We underline that, when \e{AS2} does not hold, then the residues in \e{KK44} are active, hence equations \e{HG3} and \e{HG5} still apply, but the \ac{FA} and \ac{MD} probabilities must be replaced by their complement probability counterparts.}

As we already discussed, both approximation \e{HG3} and \e{HG5} are expected to be tight for a wide parameter range, but the asymptotic expansion lacks of a measure of tightness. Top this aim, we could numerically evaluate the integral via Theorem~\ref{theoN4} or \e{KK41}, which is an option. Alternatively, the approach proposed in \cite[Theorem~7]{Erseghe15} can be used to identify on wether the chosen approximation is an upper or a lower bound to the true probability value, which in turn provides a measure of the approximation error (given by the difference between the two bounds). The rationale can be enunciated by the following result. 

\begin{theorem}\label{theoN8}
Under \e{AS2}, a sufficient condition for the validity of bounds 
$$
\eqalign{
P^{(2)}\sub{FA} & \le P\sub{FA} \le P^{(1)}\sub{FA} \cr
P^{(2)}\sub{MD} & \le P\sub{MD} \le P^{(1)}\sub{MD} \;,
}
\e{QW2}
$$ 
is that inequalities
$$
1 - g(s_x) u^2 \le s_x \sqrt{a_2}\,\Im[c_x(u)] \le 1\;,
\e{QW4}
$$
hold for $u\in[0,\infty)$, and for $x$ taking both values $\alpha$ and $\beta$. The condition is also sufficient to ensure 
$$
\eqalign{
\overline{R}^{(1)} & \le \overline{R} \le \overline{R}^{(2)} \cr
\underline{P}_e^{(2)} & \le \underline{P}_e \le \underline{P}_e^{(1)} \;,
}
\e{QW6}
$$ 
where the superscript $^{(K)}$ identifies a bound derived by use of either the $K=1$ or the $K=2$ term approximation.\hfill~$\Box$\end{theorem}
\begin{IEEEproof}
See \cite[Theorem~7]{Erseghe15}.
\end{IEEEproof}

The bounds of Theorem~\ref{theoN8} have been found to apply to all the cases of practical interest.

\subsection{Relation with the normal approximation}

Approximation \e{HG3} is useful to derive a significant property of the \ac{PPV} \Tom{meta-converse} bound, namely that it approaches capacity as $n$ grows to infinity, and, more importantly, that it is consistent with the normal approximation \e{UY2}. This provides a very neat relation with the results on parallel \ac{AWGN} channels available from the literature, and characterizes the \ac{PPV} \Tom{meta-converse} bound as an asymptotically optimum bound. Although this is already known from \cite{polyanskiy2010channel}, approximation \e{HG3} provides a very simple way to assess the result.

\begin{theorem}\label{theo8}
In a parallel \ac{AWGN} channels scenario, the \ac{PPV} \Tom{meta-converse} bound is consistent with the normal approximation \e{UY2} that uses the channel dispersion coefficient
$$
V =   \Tom{\frac1K}\sum_{k=1}^K  \frac{\snr_k(2+\snr_k)}{2(1+\snr_k)^2}\;,
\e{KK64bis}
$$
in the sense that $\overline{R}=R\sub{NA} + O(1/n)$ for $n\rightarrow\infty$.\hfill~$\Box$
\end{theorem}
\begin{IEEEproof}
See the Appendix.
\end{IEEEproof}

In any case, observe that the insights provided by the true bound can be much more relevant than those given by the normal approximation, especially for low values of $n$ where the two quantities may be significantly different.

\Tom{A summary of all the above results is available in \fig{FI2}.} \Tom{\bFig[t]{FI2}} 

\subsection{Water-filling application example}\label{wfil2}

In \fig{GA2} we illustrate the \ac{PPV} \Tom{meta-converse} bound in a multiple-carrier transmission scenario \Tom{with $128$ complex carriers, i.e., with $K=256$ (recall that each symbol is a QAM symbol in an OFDM context, hence there are two real valued symbols per each channel use)}. The channel attenuation is the one shown in \fig{GA2}(a), the noise level is assumed $\frac{N_0}2=10^{-12}\,$W/Hz, and an available power of $P\sub{tot}=0.5\,$W is considered. \Tom{The \ac{PPV} meta-converse bound is derived as the outcome of the optimization problem
$$
\eqalign{
&{\rm max}\, -\fract1n \log_2 P\sub{FA}^{(k)}(\lambda;s_\alpha, \{\snr_k\})\cr
& \wrt \lambda, s_\alpha, \snr_k= P_k/\sigma_k^2\cr
&\st \alpha'(s_\alpha; \lambda,\{\snr_k\}) = 0\cr
&\pst \fract1n \ln P\sub{MD}^{(k)}(\lambda;s_\beta=s_\alpha-\fract12, \{\snr_k\}) = \fract1n\ln(P_e)\cr
& \pst \sum_k \snr_k\sigma_k^2 = P\sub{tot}
}
\e{OPT2}
$$
which is solved by using standard routines in MatLab.}

The resulting \ac{PPV} \Tom{meta-converse} bound on rate is illustrated in \fig{GA2}(c), together with the normal approximation \e{UY2}. Performance is illustrated as a function of the code length, and the code length is expressed in number of OFDM symbols, $\Tom{n/K}$. The \ac{PPV} \Tom{meta-converse} bound was derived by application of Corollary~\ref{coro02}. The very small gap observed between approximations \e{HG3} and \e{HG5}, and the applicability of the sufficient condition of Theorem~\ref{theoN8}, ensure that the bound illustrated in figure is precise. Note the closeness between the \ac{PPV} \Tom{meta-converse} and the normal approximation values, except for small block lengths.\bFig[t]{GA2}

An illustration of the optimal power allocation is given in  \fig{GA2}(d) for the case where the block length corresponds to one OFDM, that is, $n=K$. Interestingly, in the comparison with the common water-filling result, an excess of power is loaded in ``\emph{good}'' carriers, while less power (or even no power) is loaded in ``\emph{bad}'' carriers. The gap with the common water-filling approach reduces with longer lengths $n$. Incidentally, a perfectly equivalent effect is appreciated by using the normal approximation (e.g., see \cite{park2012new}), although the resulting power allocation may be slightly different.

\Section[GA]{The \ac{AWGN} Channel}

\subsection{Scenario and notation}

The \ac{AWGN} channel is a special case of the parallel \ac{AWGN} channels scenario, where $K=1$. Because of the presence of a unique component, the index $k$ will be dropped in the following, so that $\snr$ is the reference \ac{SNR}, $P$ is the (average) power associated to transmission, $n=n_1$ is the packet length, and $\sigma^2$ is the noise variance. 

Although with a different notation, the \ac{PPV} \Tom{meta-converse} bound that we obtain from \sect{PA} correspond to the outcomes of \cite{Erseghe15}, to which we refer the interested reader for further insights and bound properties. In the following, we instead discuss the \ac{RCU} achievability bound, for which a reliable approximation can be derived by exploiting some of the methods already used in \sect{PA}.

\subsection{\ac{RCU} achievability bound}

As a first step towards our final aim, we write the result of Theorem~\ref{theo21} in a more usable and compact form, by revealing that the number of variables involved in the definition of the \ac{RCU} bound is limited. We have:

\begin{theorem}\label{theo20}
In a \ac{AWGN} channel scenario, the \ac{RCU} bound can be expressed in the form
$$
\overline{P}_e = \E{ \min(1,2^{nR}g(\B q))},\;\Tom{\hbox{with }\B q\sim\C N\Big(\snr \B e_{n}, \fract1{n}\snr\B I_{n}\Big)}\;,
\e{DS2b}
$$
where $\B e_{n}$ is a vector of length $n$ with entry one in first position and the rest set to zero, and $\B I_{n}$ is the identity matrix of order $n$, and where
$$
g(\B q)  = \P{\rule{0mm}{4mm} \|\B q\|  \eta \ge  q_{1}}\;,\qquad \eta = \frac{\tau}{\sqrt{n-1+\tau^2}}\;,
\e{DS20}
$$
with $\tau$ a t-distributed random variable of order $n-1$.~\hfill$\Box$
\end{theorem}
\begin{IEEEproof}
See the Appendix.
\end{IEEEproof}

Note that the \ac{RCU} can be also made independent of $\B q$ and solely dependent on variable $\rho=q_1/\|\B q\|$, to obtain a very compact formulation which is much more suitable for numerical evaluation since it requires a one dimensional integration. The result can be expressed as follows.

\begin{theorem}\label{theo24}
In an \ac{AWGN} channels scenario, the \ac{RCU} bound can be formalized in the form
$$
\overline{P}_e = \int_{-1}^\lambda f_\rho(a) da  + \int_\lambda^1 \frac{g(a)}{g(\lambda)} f_\rho(a) da
\e{AV6}
$$
where $f_\rho$ is the \ac{PDF} of $\rho=q_1/\|\B q\|$ with random vector $\B q$ defined in \e{DS2b}, and where  $g(a) = \P{\eta >a}$ with random variable $\eta$ defined in \e{DS20}. The value $\lambda$ is identified by the constraint $-\frac1n \log_2g(\lambda) = R$.~\hfill$\Box$ 
\end{theorem}
\begin{IEEEproof}
The proof is trivial and it is left to the reader.
\end{IEEEproof}

Incidentally, we observe that we can safely consider $\lambda>0$. As a matter of fact, since $g(0)=\frac12$ by the symmetry of the distribution of $\eta$, a negative $\lambda$ implies $R<\frac1n$, that is $M=2^{nR}<2$ which corresponds to the presence of a unique symbol for transmission, i.e., to the absence of communication.

The statistical description of $\rho$ and $\eta$ can be given in the closed form, in such a way that meaningful and compact approximations to the functions in \e{AV6} can be identified. For the statistical description of $\rho$ we can exploit approximations available from the literature for non-central t-distributed random variables (e.g., see \cite{AbramowitzStegun68}). We have:

\begin{theorem}\label{theo26}
The \ac{PDF} of random variable $\rho$ is given by
$$
\eqalign{
f_\rho(a)  & =\frac{(1-a^2)^{\frac{n-3}2}\, e^{-\frac12n(1-\frac12a^2)\snr} }{2^{\frac n2-1} \Gamma(\fract12)\Gamma(\fract{n-1}2) /\Gamma(n)}  \, U(n-\fract12;-\sqrt{n\snr} a) \cr
}
\e{AV22}
$$
with $a\in(-1,1)$, and where $U(a;x)$ is Weber's form for the parabolic cylinder function and $\Gamma(a)$ is the gamma function. The \ac{PDF} can also be written in the form $f_\rho(a) =  e^{n u_n(a)}$, where $u_n(a)$ can be approximated asymptotically by the expression
$$
u_n(a)  = u^{(0)}(a) +\frac{\ln(n)}{2n}-\frac{u^{(1)}(a)}{2n}   + O(n^{-2})\;,
\e{AV302}
$$
where
$$
\eqalign{
u^{(0)}(a) & = \fract12\ln(1-a^2)-2\alpha^2\cr
 & \qquad +(\alpha a)^2+\alpha a\sqrt{1+(\alpha a)^2}\cr
 & \qquad   + \ln(\alpha a+ \sqrt{1+(\alpha a)^2})\cr
u^{(1)}(a) & = \ln\left(1+(\alpha a)^2+ \alpha a\sqrt{1+(\alpha a)^2}\right)\cr
 & \qquad + 3\ln(1-a^2) + \ln(2\pi)\;,
}
\e{AV304}
$$
and where we used $\alpha=\sqrt{\snr/4}$.~\hfill$\Box$ 
\end{theorem}
\begin{IEEEproof}
See the Appendix.
\end{IEEEproof}

We observe that even stronger asymptotic approximations can be derived from \cite{AbramowitzStegun68}, but we verified that, in the cases of interest, the $O(n^{-2})$ approximation given by Theorem~\ref{theo26} is sufficient.

\subsection{Laplace transform expressions}

Concerning $\eta$, that is function $g(a)$, we can instead exploit the Laplace integration method of \sect{PA}. To this aim, we rewrite $g(a)$ via a Laplace dual expression, to obtain a result which is suitable for being approximated by use of the steepest descent method.

\begin{theorem}\label{theo22}
Function $g(a)=\P{\eta\ge a}$  can be written via the Laplace integral 
$$
g(a)  = \frac1{i2\pi} \int_{\mu-i\infty}^{\mu+i\infty} \frac{e^{\frac n2 \gamma(s)}}{-s} ds
\e{DS22}
$$
where $\mu<0$ and $\gamma(s)=G_{\frac{n}2}(s)+as$, and where the function $G_\nu$ is defined as
$$
\eqalign{
G_\nu(s) & = \fract1\nu \ln\Big(\rule{0mm}{1em}_0F_1(;\nu; (\fract12\nu s)^2)\Big)\cr
 & = \fract1\nu \ln\left( \Gamma(\nu) \cdot  (\fract12\nu s)^{1-\nu} I_{\nu-1}(\nu s) \right)
 }
\e{CR2}
$$
with $_0F_1$ the confluent hypergeometric limit function, and $I_\nu$ the modified Bessel function of the first kind.~\hfill$\Box$
\end{theorem}
\begin{IEEEproof}
See the Appendix.
\end{IEEEproof}

The integral can be further simplified by exploiting a steepest descent path. The idea and the procedure are perfectly identical to the \ac{MD} probability expression derived in \sect{PA}, and provide (compare to \e{KK40})
$$
g(a) = 1(s_\gamma) - \frac1{i2\pi} e^{\frac n2\gamma(s_\gamma)} \int_{-\infty}^{+\infty} e^{-\frac n2 u^2} c_\gamma(u) du
\e{KU2}
$$
where $p_\gamma(u)$ is the steepest descent path which corresponds to choice $\Im[\gamma(s)]=0$, and which transits through the real valued saddle point 
$$
s_\gamma  = \argmin_{s\in\M R} \gamma(s)\;.
\e{KU4}
$$
Function $c_\gamma$ is instead defined as (compare to \e{KK40a})
$$
c_\gamma(u) = \frac{p'_\gamma(u)}{p_\gamma(u)} = \frac{2u}{-\gamma'(p_\gamma(u))p_\gamma(u)}\;.
\e{KU6}
$$
Observe from \e{DS2b} that we are interested in values $g(a)\le2^{-nR}$, i.e., we are interested in the tails of the considered distributions. Therefore we can safely assume $s_\gamma<0$, and obtain an asymptotic approximation equivalent to that of Corollary~\ref{coro02}. Being the derivation of the result perfectly identical to the steps leading to \e{HG3}, the theorem is given without proof.

\begin{theorem}\label{theo23}
Function $g(a)=  e^{-n v_n(a)}$ can be asymptomatically approximated via the expression
$$
\eqalign{
v_n(a) & = -\fract12\Big[G_{\frac n2}(s_\gamma(a))+a\,s_\gamma(a)\Big] \cr
 & \qquad + \frac1{2n}\ln\Big(\pi n s_\gamma^2(a)\,G_{\frac n2}''(s_\gamma(a)) \Big) + O(n^{-2})\;,
}
\e{AV10}
$$
where $G_\nu$ is defined in \e{CR2}, and where $s_\gamma(a)=[G_{\frac n2}']^{-1}(-a)$ is assumed to be a negative value, $s_\gamma(a)<0$.~\hfill$\Box$
\end{theorem}
\begin{IEEEproof}
The proof is left to the reader.
\end{IEEEproof}

To be usable, a $O(n^{-2})$ approximation is needed for $G_{\frac n2}(a)$ and its derivatives in Theorem~\ref{theo23}. This can be obtained from standard asymptotic expansions for the modified Bessel functions of the first kind (e.g., see \cite{AbramowitzStegun68}), and provides the following result.

\begin{theorem}\label{theoq2}
Function $G_\nu$ in \e{CR2} can be asymptotically approximated by expression
$$
\eqalign{
G_\nu(s) & =  \sqrt{1+s^2} -1 - \fract1{2\nu}\ln( \sqrt{1+s^2})  \cr
 & \qquad -\Big(1-\fract1\nu\Big)\ln\Big(\frac{1+\sqrt{1+s^2}}2\Big)  + O(1/\nu^2)\;.
}
\e{CR10}
$$~\hfill$\Box$
\end{theorem}
\begin{IEEEproof}
See the Appendix.
\end{IEEEproof}

Note that \e{CR10} reveals that
$$
G_\infty(s) = \sqrt{1+s^2}-1 -\ln\Big(\frac{1+\sqrt{1+s^2}}2\Big) \;.
\e{CR12}
$$
The rapid convergence to the limit value \e{CR12} is illustrated in \fig{GA10}, displaying the function $G_\nu(x)+\rho x$ for $\rho=0.5$, which is the core function used in the definition of $\gamma$. The behavior is equivalent for any $0<\rho<1$.\Fig[h]{GA10} 

The $O(n^{-2})$ approximation provided by \e{AV10} was verified to be very precise. Alternatively, a weaker $O(n^{-1})$ result can be obtained by exploiting \e{CR12} in Theorem~\ref{theo23}, to have, with a little effort, $s_\gamma(a)=-2a/(1-a^2)+O(n^{-1})$ and
$$
v_n(a)  =   -\fract12\ln (1-a^2)  + \frac{\ln(n)}{2n} +O(n^{-1})\;.
\e{AV10b}
$$

\subsection{Reliable approximation}

With the notation assessed above the \ac{RCU} bound \e{AV6} assumes the form
$$
\eqalign{
\overline{P}_e & = \int_{-1}^\lambda e^{n u_n(a)} da +  \int_\lambda^1 e^{n [u_n(a) - v_n(a) + R\ln(2)]} da
}
\e{AV32}
$$
where $\lambda$ is defined by the relation $v_n(\lambda)=R\ln(2)$. We also observe from \e{AV10b} that
$$
\lambda = \sqrt{1-2^{-2R}\,e^{\ln(n)/n}} + O(n^{-1})
\e{AV14bi}
$$
holds for $\lambda$. 

Integral \e{AV32} can be approached numerically, which is an option, but a very compact and analytical result is obtained by reliably approximating it by use of an approach similar to the Laplace's method \cite[\S5]{Bleistein75} (which exploits a quadratic approximation of the exponent at its maximum value). In particular, since in the considered context the maximum value lies outside  of the integration region,  a linear approximation of the exponent will be exploited. Hence, we obtain the following result.

\begin{theorem}\label{theo28}
In a \ac{AWGN} channel scenario, for sufficiently large $n$ the \ac{RCU} bound can be approximated in the form
$$
\frac1n\ln \overline{P}_e  = u^{(0)}(\lambda) - \frac{\ln(2\pi n)}{2n} -\frac1{n} \ln(w^{(0)}(\lambda))+O(n^{-2})
\e{AV33}
$$
where
$$
\eqalign{
w^{(0)}(a) & = \sqrt{\frac{(1-a^2)}{a^2}\cdot \Big(1+\alpha a w^{(1)}(a)\Big)}\cr
 & \qquad \cdot  \Big(2\alpha (1-a^2) w^{(1)}(a) - a\Big)\cr
 & \qquad \cdot\Big(2a-2\alpha (1-a^2) w^{(1)}(a) \Big)\cr
w^{(1)}(a) & = \alpha a+\sqrt{1+(\alpha a)^2}\;,\qquad \alpha=\sqrt{\fract14\snr}\;,
}
\e{AV33c}
$$
and where $\lambda$ is defined by the equivalence $v_n(\lambda)=R\ln(2)$ with $v_n$ expressed by the $O(n^{-2})$ approximation \e{AV10} using \e{CR10}. For a correct applicability of the theorem, $w^{(0)}(\lambda)$ must be a positive value. 
~\hfill$\Box$ 
\end{theorem}
\begin{IEEEproof}
See the Appendix.
\end{IEEEproof}

As discussed in the proof, the request $w^{(0)}(\lambda)>0$ practically corresponds to requiring small error probabilities, $P_e<\frac12$, and does not limit the applicability of the result. \Tom{Incidentally, $O(n^{-3})$ results can also be derived with some additional effort, by deriving more refined asymptotic expressions for $u_n$ and $v_n$, and by subsequently applying the rationale of \cite{tierney1986accurate}.}

The asymptotic expression given by Theorem~\ref{theo28} further allows to easily verify the validity of the normal approximation for the \ac{RCU} bound, thus providing an alternative derivation to the result of \cite{tan2015it}.

\begin{theorem}\label{theo28b}
In a \ac{AWGN} channel scenario, the \ac{RCU} achievability bound is consistent with the normal approximation \e{UY2} that uses the channel dispersion coefficient
$$
V = \frac{\snr(2+\snr)}{2(1+\snr)^2}\;,
\e{KK64bis2}
$$
in the sense that $\underline{R}=R\sub{NA} + O(1/n)$ for $n\rightarrow\infty$.\hfill~$\Box$
\end{theorem}
\begin{IEEEproof}
See the Appendix.
\end{IEEEproof}

\Tom{A summary of all the above results, valid for the \ac{AWGN} channel scenario, is given in \fig{FI4}, where we also explicitly recall that the \ac{RCU} bound can be interpreted both as a bound on error probability, as well as a bound on rate (see the paragraph after Theorem~\ref{theo21}).}\Tom{\bFig[t]{FI4}}

\subsection{Numerical examples}

A few examples of application of the \ac{PPV} \Tom{meta-converse} bound that uses Corollary~\ref{coro02} with $K=1$, and of the \ac{RCU} bound that uses Theorem~\ref{theo28}, are shown in \fig{GA4}. In \fig{GA4}.(a) and (b) we illustrate the scenarios depicted in \cite[Fig.~6-7, 12-13]{Polyanskiy10}, with the major difference that the normal approximation in \cite{Polyanskiy10} is neglecting the $\log_2(n)/(2n)$ term in \e{UY2}, hence it is less accurate. For reasons of space/readability, the Shannon bounds \cite{Shannon59} are not shown. These are, however, very close to the \ac{PPV} \Tom{meta-converse} and the \ac{RCU} bounds, the \ac{RCU} bound slightly improving the achievability limit. 

The closeness between the upper \ac{PPV} \Tom{meta-converse} and the lower \ac{RCU} bounds can be further appreciated in \fig{GA4}.(c), which provides a wider look onto the spectral efficiency $\rho=2R$ versus the $E_b/N_0$ measure for quite a large range of blocklengths $n$. Plots are given for those regions where the discussed $O(n^{-2})$ approximations provide a reliable result. Note in figure how the normal approximation \e{UY2} provides a very good fit. However, as illustrated in \fig{GA4}.(d) under an error probability perspective, the normal approximation might be loose for low block lengths, where the \ac{RCU} bound occurs at a (non negligible) $0.1$-$0.2\,$dB distance in \ac{SNR}. 

Overall, the indications we get from \fig{GA4} is that the normal approximation  (for both achievability and converse bounds) is meaningful and simple with moderate to large blocklengths, while for short blocklengths the \ac{PPV} \Tom{meta-converse} and the \ac{RCU} bounds provide a reliable yet sufficiently simple alternative. 
\bFig[!p]{GA4}

\Section[SO]{Binary-Input \ac{AWGN} Channel}

\subsection{Scenario and notation}

Binary coding under an \ac{AWGN} channel, also known as the BI-\ac{AWGN} channel, is a scenario that corresponds to a soft-decoding receiver implementation, and which sets the limits for the (classical) performance of a binary code. In this case the codewords set of interest is 
$$
\C K = \{1,-1\}^n\;,
\e{SD2}
$$
and the channel \ac{PDF} is a standard \ac{AWGN} expression which we write in the form
$$
p_{y|x}(\B b|\B a) = \left(\frac\snr{2\pi}\right)^{\frac n2} e^{-\frac1{2}\snr \|\Bs b-\Bs a\|^2}\;,
\e{SD4}
$$
where $\snr=P/\sigma^2$ is the reference \ac{SNR} value. Capacity is achieved with equally likely input symbols, $p_x(\B a) =  2^{-n}$, and guarantees an output \ac{PDF}
$$
p_y(\B b) =  \prod_{i=1}^n \sqrt{\frac\snr{2\pi}}  \frac{e^{-\frac1{2}\snr (b_i-1)^2}  + e^{-\frac1{2}\snr (b_i+1)^2} }2\;,
\e{SD8}
$$
which provides a capacity expression of the form
$$
C = 1 - \sqrt{\frac\snr{2\pi}} \int  e^{-\frac12\snr(b-1)^2} \log_2(1+e^{-2\snr b})\;db\;.
\e{SD9}
$$

\subsection{\ac{PPV} \Tom{meta-converse} bound}

The \ac{PPV} \Tom{meta-converse} bound can be derived by using the same techniques employed in the parallel \ac{AWGN} channel case, that is through a representation via a Lagrange integral and by its asymptotic approximation. As a starting point, we observe that the compact expression for the \ac{FA} and \ac{MD} probabilities is independent of the specific codeword choice, ant that Theorem~\ref{theo1} can be used in the following form.

\begin{theorem}\label{sdTEO1}
In a \ac{AWGN} channel scenario with binary codewords the \ac{FA} and \ac{MD} probabilities \e{KK12} can be expressed by
$$
\eqalign{
P\sub{FA}(\lambda)& = \P{\sum_{i=1}^n h(u_i) \le n\lambda}\cr
P\sub{MD}(\lambda) & =  \P{\sum_{i=1}^n h(v_i) > n\lambda }\cr
}
\e{SD20}
$$
where
$$
\Tom{\eqalign{
u_i& \sim\C N(d_i\snr,\snr)\;,\quad p_{d_i}(1)=p_{d_i}(-1)=\fract12\cr
v_i& \sim \C N(\snr,\snr)
}}
\e{SD20sec}
$$
and  $h(x)=\ln(1+e^{-2x})$.\hfill~$\Box$
\afterpage{\clearpage}
\end{theorem}
\begin{IEEEproof}
See the Appendix.
\end{IEEEproof}

The Laplace transform method of Theorem~\ref{theo3} can then be exploited to write the result in an equivalent form. The proof is left to the reader, since it is a simple re-application of the concepts that led to Theorem~\ref{theo3}.

\begin{theorem}\label{sdTEO2}
In a \ac{AWGN} channel scenario with binary codewords the \ac{FA} and \ac{MD} probabilities \e{KK12} can be expressed as in \e{KK30} where
$$
\eqalign{
\alpha(s) &=  \beta(s-1) -2\ln(2) + 2\lambda\cr
\beta(s) &= 2 \Big(\lambda s+\ln(H(s))\Big)
}
\e{SD30}
$$
and where 
$$
H(s) =  \frac1{\sqrt{2\pi \snr}} \int_{-\infty}^\infty  e^{-\frac1{2\snr} (x-\snr)^2} e^{-s \cdot h(x)} dx
\e{SD32}
$$
is a Laplace transform which converges for any $s\in\M C$.\hfill~$\Box$
\end{theorem}
\begin{IEEEproof}
The proof is left to the reader.
\end{IEEEproof} 

Incidentally, by exploiting the above notation we can denote capacity in the form
$$
C = 1+ \frac{H'(0)}{\ln(2)}\;,
\e{SD22}
$$
where $H'$ is the derivative of \e{SD32} with respect to $s$.

The main difficulty involved with Theorem~\ref{sdTEO2} is to identify a usable analytic expression for \e{SD32}, which is hardly obtainable. In any case, numerical evaluation can lead to the identification of saddle points, and then steepest descent paths for numerical integration in the form of Theorem~\ref{theoN4}. A further option of interest is to exploit an asymptotic approximation in the form discussed in \Tom{Procedure~\ref{theo5}}, which leads to the compact expressions  \e{HG3} and \e{HG5}. The result is in this case a mixture of analytical expressions and numerical integrations.

\begin{procedura}\label{sdTEO3}
In a \ac{AWGN} channel scenario with binary codewords the \ac{FA} and \ac{MD} probabilities \e{KK12} allow the asymptotic expansion \Tom{\e{KK44} whose coefficients can be evaluated according to Procedure~\ref{theo5} with the following substitutions}: 
\begin{itemize}
\item[1)] the functions $\alpha$ and $\beta$ \Tom{must be} defined as in \e{SD30}; 
\item[2)] $s_\alpha$ and $s_\beta$  \Tom{must be} derived, respectively, from equivalences $\alpha'(s_\alpha)=0$ and $\beta'(s_\beta)=0$, to satisfy $s_\alpha=s_\beta+1$; 
\item[3)] the coefficients $a_k$  \Tom{must be} defined as $a_k=\beta^{(k)}(s_\beta)/k!$, with $\beta^{(k)}$ the $k$th derivative of the function $\beta$ in \e{SD30}. 
\end{itemize}
In this context, the $O(n^{-2})$ and $O(n^{-3})$ approximations given, respectively, by \e{HG3} and \e{HG5}-\e{HG7} are valid under the assumption that $s_\alpha>0$, and $s_\beta<0$.\hfill~$\Box$
\end{procedura}
\begin{IEEEproof}
The proof is identical to the one of \Tom{Procedure~\ref{theo5}}, and it is left to the reader.
\end{IEEEproof}

We observe that, by using notation
$$
H^{(\ell)}(s) =  \frac1{\sqrt{2\pi \snr}} \int_{-\infty}^\infty  e^{-\frac1{2\snr} (x-\snr)^2} e^{-s \cdot h(x)} (-h(x))^\ell\, dx\;,
\e{SD32b}
$$
for the derivatives of $H$, the first few derivatives of $\beta$ are given by
$$
\eqalign{
\beta'(s) & = 2\left(\lambda + \frac{H^{(1)}(s)}{H(s)} \right)\cr
\beta''(s) & = 2 \left(\frac{H^{(2)}(s)}{H(s)}-\left( \frac{H^{(1)}(s)}{H(s)}\right)^2\right)\cr
\beta'''(s) & = 2 \left(\frac{H'''(s)}{H(s)}-3\frac{H''(s)H'(s)}{H^2(s)}+2\left( \frac{H'(s)}{H(s)}\right)^3\right)\cr
\beta^{iv}(s) & = 2 \left(\frac{H^{iv}(s)}{H(s)}-4\frac{H'''(s)H'(s)}{H^2(s)}-3\left( \frac{H''(s)}{H(s)}\right)^2\right.\cr
 & \hspace*{16mm} \left.+12\frac{H''(s)(H'(s))^2}{H^3(s)}-6\left( \frac{H'(s)}{H(s)}\right)^4\right)\;.
}
\e{SD52}
$$

A certificate on the  approximation error (e.g., in the form of Theorem~\ref{theoN8}) is difficult to obtain in the binary codewords scenario. Nevertheless, we can always resort to Theorem~\ref{theoN4} in order to identify a robust numerical integration method. A somewhat weaker (but much simpler) guarantee is obtained by comparing $O(n^{-2})$ and $O(n^{-3})$ approximations of \Tom{Procedure~\ref{sdTEO3}}, and by validating the result only in case of agreement. Validity of the normal approximation is instead ensured in the following form.

\begin{theorem}\label{theo8bi}
In a \ac{AWGN} channel scenario with binary codewords,  the \ac{PPV} \Tom{meta-converse}  bound \e{KK14} of Theorem~\ref{theo1} is consistent with the normal approximation \e{UY2} that uses the channel dispersion coefficient
$$
V = \fract12\beta''(0)\;,
\e{NAA2}
$$
that is, $\overline{R}=R\sub{NA} + O(1/n)$ for $n\rightarrow\infty$. \hfill~$\Box$
\end{theorem}
\begin{IEEEproof}[Proofs]
See the Appendix.
\end{IEEEproof}

\Tom{A summary of all the above results on the BI-\ac{AWGN} channel scenario is given in compact procedural form in \fig{FI6}.} 
\Tom{\bFig[t]{FI6}} 

\subsection{$\kappa\beta$ achievability bound}

In the \ac{AWGN} channel scenario with binary codewords, the \ac{RCU} bound can be compactly expressed in the following form.

\begin{theorem}\label{theo40pre}
In a \ac{AWGN} channel scenario with binary codewords, the \ac{RCU} bound can be expressed as
$$
\overline{P}_e =  \E{ \min(1,2^{nR}g(\B w))},\; \Tom{\hbox{with }\B w\sim\C N(\B 1\snr,\B I\snr)}
\e{SDD20}
$$
and where
$$
g(\B w) = \P{ \sum_{i=1}^n d_iw_i \le 0},\; \Tom{\hbox{with } p_{d_i}(0)=p_{d_i}(1)=\fract12 }\;.
\e{SDD21}
$$
~\hfill$\Box$
\end{theorem}
\begin{IEEEproof}
See the Appendix.
\end{IEEEproof}

Although it has a very simple expression, the function $g(\B w)$ can hardly be mapped into a mono-dimensional integral, and the approach used for \e{DS20} (which easily mapped into \e{AV6}) is not applicable. As a consequence, the integration region of interest assumes a composite multidimensional form, which is hardly usable to obtain a satisfactory result. We therefore investigate the (weaker) $\kappa\beta$ bound which, in this context, assumes a very simple form.

\begin{theorem}\label{theo40}
In a \ac{AWGN} channel scenario with binary codewords, the $\kappa\beta$ bound \e{KB2} uses $\kappa(\tau)=\tau$, so that the achievability bound can be expressed in the form
$$
 \underline{R} = \max_{\alpha\in(0,1)} \fract1n\log_2((1-\alpha)P_e) - \fract1n\log_2( P\sub{FA}(\lambda))
\e{KB12}
$$
where $\lambda$ is set by the constraint $P\sub{MD}(\lambda) = \alpha P_e$. The achievability bound on error probability instead reads as
$$
\overline{P}_e = \min_{\alpha\in(0,1)} \frac{P\sub{MD}(\lambda)}{\alpha} \;,
\e{KB14}
$$
where $\lambda$ is set by the constraint $R = \frac1n\log_2((1-\alpha) P\sub{MD}(\lambda))-\frac1n\log_2(\alpha P\sub{FA}(\lambda))$.~\hfill$\Box$
\end{theorem}
\begin{IEEEproof}
See the Appendix.
\end{IEEEproof}

The $\kappa\beta$ bound can be readily evaluated (or approximated) by using the results of Theorem~\ref{sdTEO2} and \Tom{Procedure~\ref{sdTEO3}}. Also, its asymptotic behavior is easily obtained from Theorem~\ref{theo8bi}, which ensures that for a fixed $\alpha$ the bound behaves as
$$
\eqalign{
\underline{R}_\alpha & = C  - \sqrt{\frac Vn}\,\log_2(e)\,Q^{-1}(\alpha P_e) + \fract1n\log_2(1-\alpha)\cr
 & \qquad + \frac{\log_2(n)}{2n} + O(n^{-1})\;.
}
\e{KB14b}
$$
The optimum approximation $\alpha=1-O(n^{-\frac12})$ further ensures the (weak) relation $\underline{R}=R\sub{NA} + O(\log_2(n)/n)$, which was already observed in \cite{Polyanskiy10} in the (somewhat similar) \ac{AWGN} channel case.

\subsection{Numerical examples}

An overview of the \ac{PPV} \Tom{meta-converse}  and achievability bounds for the \ac{AWGN} channel with binary coding is given in \fig{GA6}. The plots of \fig{GA6}.(a) and (b) are, respectively, the counterparts to the plots of  \fig{GA4}.(c) and (d), in such a way to provide a compact view on the characteristics of the converse and achievability  bounds, and of their relation with the normal approximation. The kind of considerations which can be drawn are equivalent to the ones already discussed in the \ac{AWGN} case, to which we refer, with the addition of the fact that the $\kappa\beta$ bound is evidently a not a strikingly tight bound. See also \cite{Erseghe15} for the comparison with the $\kappa\beta$ bound in the \ac{AWGN} case.
\bFig[!p]{GA6}
\afterpage{\clearpage} 

Some further aspects are covered in \fig{GA6}.(c) and (d). 

In \fig{GA6}.(c) we show  the significant improvement given by the \ac{PPV} \Tom{meta-converse} bound with respect to the \ac{ISP} bound of \cite{wiechman:it08}. The \ac{ISP} bound is the state-of-the-art for converse bounds in an \ac{AWGN} channel with binary coding, and it is an improvement over the bounds by Valembois and Fossorier \cite{Valembois:it04}, as well as over over Shannon, Gallager, and Berlekamp's 1967 bound  \cite{SHANNON1967}. Incidentally, the codes shown in \fig{GA6}.(c) are those of \cite[Fig.~2,3, and 4]{wiechman:it08}, to which the reader is referred. The $\kappa\beta$ bound, which is not shown in \fig{GA6}.(c),  provides a performance roughly equivalent to those of the achievability results illustrated in \cite{wiechman:it08}.

Finally \fig{GA6}.(d) shows the existing $1\,$dB gap between the \ac{PPV} \Tom{meta-converse} (or $\kappa\beta$) bound and the practical performance of \ac{LDPC} decoding via belief propagation (i.e., message passing). The codes used in figure are both rate $R=\frac12$, the shorter one being the $(k,n)=(1320,2640)$ Margulis code \cite{Margulis:co82}, the longer one being taken from \cite{johnson2009iterative}.  \Tom{It is in any case worth recalling that, for very short packet sizes one can construct codes (and decoders) that approach the achievability bound, and in some regions (high error rate) even beat it \cite{liva2013short}.}

\Section[HA]{The Binary Symmetric Channel}

\subsection{Scenario and notation}

The \ac{BSC} is a numerical scenario where input and output channel symbols are binary, $\B x,\B y\in\{\pm1\}^n$, and the transition probabilities are fully described by the crossover probability $P\sub{bit}$, where we assume with no loss in generality that $0<P\sub{bit}<\frac12$. It covers the hard-decoder implementation under binary transmission and \ac{AWGN} channel, provided that the incorrect transition probability is defined as
$$
P\sub{bit} = Q(\sqrt{\Gamma})
\e{HD2}
$$
where $Q(\cdot)$ is the Gaussian \ac{CCDF}. The codewords set of interest is, equivalently to the soft-decision counterpart \e{SD2}, $\C K = \{1,-1\}^n$, and the channel transition probabilities take the form
$$
p_{y|x}(\B b|\B a)  = (1-P\sub{bit})^n\,\left(\frac{P\sub{bit}}{1-P\sub{bit}}\right)^{\|\Bs b-\Bs a\|_H}
\e{HD4}
$$
with $\|\cdot\|_H$ denoting the Hamming weight. Capacity is achieved for equally likely input symbols, $p_x(\B a) =  2^{-n}$, which guarantees that also the output symbols are equally likely, that is, $p_y(\B b) =  2^{-n}$. The closed-form capacity expression is
$$
C = 1-h(P\sub{bit}) 
\e{HD10}
$$
where
$$
h(x) = -x\log_2x-(1-x)\log_2(1-x)
\e{HD12}
$$
is the binary random variable entropy function.

\subsection{\ac{PPV} \Tom{meta-converse} bound}

For a \ac{BSC}, the \ac{PPV} \Tom{meta-converse}  bound is expressed by the following result which takes into account the appropriate modifications to Theorem~\ref{theo1} needed for discrete random variables.

\begin{theorem}\label{hdTEO1}
In a \ac{BSC} channel scenario the \ac{PPV} \Tom{meta-converse} bound can be expressed in the form
$$
R \le \overline{R} = -\fract1n\log_2\Big( P\sub{FA}(d) -\zeta\,q_n(d;\fract12) \Big)\;,
\e{ADO2}
$$
where the integer constant $0\le d\le n$ and the real valued constant $0\le\zeta< 1$ are defined by 
$$
P_e = P\sub{MD}(d) + \zeta q_n(d;P\sub{bit})\;,
\e{ADO2cc}
$$
and where we used
$$
q_n(d;\epsilon) = \left({n\atop d}\right) (1-\epsilon)^{n-d}\,\epsilon^d\;.
\e{ADO2bb}
$$
In the above context, the \ac{FA} and \ac{MD} probabilities \e{KK12} take the form
$$
\eqalign{
P\sub{FA}(d) & = \P{\sum_{i=1}^n u_i\le  d}\cr
P\sub{MD}(d) & = \P{\sum_{i=1}^n v_i> d}\;,
}
\e{HD20}
$$
where
$$
\Tom{\eqalign{
p_{u_i}(0)&=p_{u_i}(1)=\fract12\cr
p_{v_i}(0)&=1\!-\!P\sub{bit}\;,\quad p_{v_i}(1)=P\sub{bit}\;.
}}
$$
Operatively, the value of $d$ corresponds to the minimum $d$ which satisfies $P\sub{MD}(d)\le P_e$. \hfill~$\Box$
\end{theorem}
\begin{IEEEproof}
See the Appendix.
\end{IEEEproof}

Observe that the \ac{FA} probability in \e{HD20} is a binomial \ac{CDF} of order $n$ and parameter $\frac12$, and the \ac{MD} probability is a binomial \ac{CCDF} of parameter $P\sub{bit}$. The result of Theorem~\ref{hdTEO1} is also identical to \cite[Theorem~35]{Polyanskiy10}, where binomial \ac{CDF} and \ac{CCDF} are implicitly given in terms of a recurrence relation. A lower bound $\underline{P}_e$ can also be derived from Theorem~\ref{hdTEO1} by setting $\underline{P}_e=P\sub{MD}(d) + \zeta q_n(d;P\sub{bit})$ and using constraint $P\sub{FA}(d) -\zeta\,q_n(d;\fract12)=2^{-nR}$. In this case the value of $d$ corresponds to the minimum $d$ which satisfies $P\sub{FA}(d)\ge 2^{-nR}$.

The result of Theorem~\ref{hdTEO1} calls for an asymptotic expression of the binomial \ac{CDF} for large values of $n$, since the standard series expansion (e.g., the above cited recurrence relation) may be hardly applicable for large values of $n$. Other results available from the literature, e.g., the asymptotic expansion given in \cite{Temme75}, are not applicable in the present context. Use of  Hoeffding's theorem \cite[Theorem~1]{Hoeffding63} is also a widely used option for approximating binomial \acp{CDF} but, although able to capture the limit behavior of the bound, it is very weak and in general unusable. We therefore follow the Laplace domain approach, and look for an asymptotic expansion which holds uniformly for large $n$. The starting point is provided by the following result.

\begin{theorem}\label{hdTEO2}
In a \ac{BSC} channel scenario the \ac{FA} and \ac{MD} probabilities \e{HD20} can be expressed, for $d$ not an integer, as in \e{KK30} where
$$
\eqalign{
\alpha(s) & = 2\lambda s+2\ln(1+e^{-s})-2\ln(2) \cr
\beta(s) & = \alpha(s+\delta_0) -2(\lambda-P\sub{bit})\delta_0 + 2\ln(2)\cdot C\;,
}
\e{HD32}
$$
with 
$$
\lambda = \frac{d}{n}\;,\qquad \delta_0  = \ln((1-P\sub{bit})/P\sub{bit})\;, 
\e{HD32b}
$$
and $\delta_0 >0$. Values for integer $d$ can be derived from left and right limits.\hfill~$\Box$
\end{theorem}
\begin{IEEEproof}
See the Appendix.
\end{IEEEproof}

Note that in Theorem~\ref{hdTEO2} we assume $d$ not an integer in order to avoid the points of discontinuity where the inverse Laplace transform equals the average of the left and right limits. We also observe that, being variables integer valued, a perfectly equivalent result could be obtained by use of the z-transform. However, we keep the Laplace notation for coherence with the results developed so far. The final result, as well as the intermediate steps, will be the same.

\subsection{On the correct identification of the steepest descent path}

Although Theorem~\ref{hdTEO2} carries the same structure of Theorem~\ref{theo3}, integration via the method of steepest descent is more involved since, because of the discrete nature of the output, the integration path is now a collection of integration paths. As a matter of fact, the functions $\alpha$ and $\beta$ carry a periodic behavior. Besides the real valued saddle points
$$
s_\alpha = \ln \left( \frac{1-\lambda}{\lambda} \right)\;,\quad
s_\beta = s_\alpha -\delta_0\;,
\e{HD34}
$$
any point of the form $s_\alpha+i2k\pi$ and $s_\beta+i2k\pi$, for any choice of $k\in\M Z$, is a saddle point of, respectively, $\alpha$ or $\beta$. The steepest descent paths associated with \e{HD34} must then be connected as graphically illustrated in \fig{SH10}. \Fig[h]{SH10}

The identification of the arc passing through saddle point $s_\alpha$ can be performed by investigating the equivalence $\Im(\alpha(s))=0$ for $\Im(s)=\phi\in(-\pi,\pi)$, ad it provides the expression 
$$
\C A_\alpha = \Big\{p_\alpha(\phi)=\ln(v(\phi))+ i\phi, \phi\in(-\pi,\pi)\Big\}\;,
\e{HD37b}
$$ 
where
$$
v(\phi) = \frac{\sin ( (1-\lambda)\phi)}{\sin(\lambda \phi )}\;.
\e{HD37}
$$
The arc $\C A_\alpha$ therefore links the points $-i\pi$ and $i\pi$, and transits through $s_\alpha$, which corresponds to the choice $\phi=0$. The integration paths then assume the form 
$$
\C L_\alpha=\bigcup_{k=-\infty}^{+\infty} \{\C A_\alpha+i2k\pi\}\;, 
\e{HD37c}
$$
and $\C L_\beta=\C L_\alpha-\delta_0$, which, with some effort, provides the following result. 

\begin{theorem}\label{hdTEO2bis}
In a \ac{BSC} channel scenario the \ac{FA} and \ac{MD} probabilities \e{HD20} can be expressed, for $d$ an integer, in the form
$$
\eqalign{
P\sub{FA}(d) & = 1(-s_\alpha)+ \frac1{i2\pi} \int_{-\pi}^{\pi} \frac{e^{\frac n2\alpha(p_\alpha(\phi))} }{1-e^{-p_\alpha(\phi)}} p'_\alpha(\phi)\;d\phi\cr
P\sub{MD}(d) & = 1(s_\beta)-\frac1{i2\pi} \int_{-\pi}^{\pi} \frac{e^{\frac n2\beta(p_\beta(\phi))} }{1-e^{-p_\beta(\phi)}} p'_\beta(\phi)\;d\phi\cr
}
\e{HDW2}
$$
where constants and functions are defined in \e{HD32}-\e{HD37}, and $p_\beta(\phi)=p_\alpha(\phi)-\delta_0$. \hfill~$\Box$
\end{theorem}
\begin{IEEEproof}
See the Appendix.
\end{IEEEproof}

Observe that, in \e{HDW2}, it is 
$$
\eqalign{
\alpha(p_\alpha(\phi))  & = 2 \ln (2)\cdot  h(\lambda)  - 2 d(\phi) - 2 \ln( 2)\cr
 & = \alpha(s_\alpha)- 2 d(\phi)\cr
\beta(p_\beta(\phi))  & = 2 \ln (2) \cdot h(\lambda) - 2 d(\phi) \cr
& \qquad +2\Big[\lambda\ln(P\sub{bit})+(1-\lambda)\ln(1-P\sub{bit}) \big] \cr
& =  \beta(s_\beta)- 2 d(\phi)\cr
 }
\e{HD39}
$$
where, by exploiting (with a little effort) standard trigonometric identities, we can derive that
$$
\eqalign{
d(\phi) 
 &= \lambda \ln\sinc(\lambda\phi) + (1-\lambda)\ln\sinc((1-\lambda)\phi)\cr
 & \hspace*{28mm}-  \ln\sinc(\phi)
}
\e{HD38}
$$
with $\sinc(x) =\sin(x)/x$. In $\phi\in(-\pi,\pi)$  the function $d(\phi)$ is a real valued, positive, convex function with minimum value $d(0)=0$, and with $d(\pm\pi)=\infty$. \Tom{A summary of the resulting integral expression for the \ac{FA} and \ac{MD} probabilities is available in compact explicit form in \fig{FI8}.}

The above also allows obtaining a form equivalent to Theorem~\ref{theoN4} by introducing variable $u$ in the relation $\alpha(p_\alpha(\phi))=\alpha(s_\alpha)-u^2$, that is, $u^2=2d(\phi)$. We obtain the following result.

\begin{theorem}\label{hdTEO2ter}
In a \ac{BSC} channel scenario the \ac{FA} and \ac{MD} probabilities \e{HD20} can be expressed, for $d$ an integer, in the form \e{KK40} of Theorem~\ref{theoN4} where 
$$
\eqalign{
c_x(u)  & = \frac{p'_x(u)}{1-e^{-p_x(u)}} = \frac{2u}{-\alpha'(p_\alpha(u))\,(1-e^{-p_x(u)})} \cr  
	   &= -c^*_x(-u)\cr
p_\alpha(u) &=p_\alpha(\phi(u))\cr
\phi(u) & =f^{-1}(u)\;,\quad f(\phi) = {\rm sign}(\phi)\sqrt{2 d(\phi)}\;,
}
\e{KK40a2}
$$
where $p_\beta(\phi)=p_\alpha(\phi)-\delta_0$, and where the remaining constants and functions are defined as in \e{HD32}-\e{HD37}. \hfill~$\Box$
\end{theorem}
\begin{IEEEproof}
The proof is left to the reader.
\end{IEEEproof}

\subsection{Asymptotic expansion}

The result obtained implies that the asymptotic expansion of \Tom{Procedure~\ref{theo5}} holds, but a different strategy must be used for evaluating the coefficients by standard algebraic operations on Taylor series. The compact result is the following. Details are tedious and are therefore skipped, but the method has been already discussed in the proof of  \Tom{Procedure~\ref{theo5}}.

\begin{procedura}\label{hdTEO5}
In a \ac{BSC} channel scenario the \ac{FA} and \ac{MD} probabilities \e{HD20} allow the asymptotic expansion \e{KK44} of \Tom{Procedure~\ref{theo5}} where $c_{\alpha,k}$ and $c_{\beta,k}$ are the Taylor expansion coefficients of functions $c_\alpha$ and $c_\beta$, respectively, as given in Theorem~\ref{hdTEO2ter}. Operatively, the following iterative procedure can be used to identify the coefficients, provided that, at step 4), $q$ is set to $q=0$ for the \ac{FA} probabilities, and to $q=\delta_0$ for the \ac{MD} probabilities.
\begin{enumerate}
\item Evaluate the real valued Taylor series coefficients in $\phi=0$ of the function $d(\phi)$ in \e{HD38} \Tom{according to rule:} $d_{2k+1}=0$, $d_0=0$, and 
$$
d_{2k} = \frac{ (-1)^{k+1} 2^{2k-1} B_{2k}}{(2k)!\cdot k} [1-\lambda^{2k+1}-(1-\lambda)^{2k+1}]
\e{HD50a}
$$
where $B_{2k}$ is the Bernoulli number \cite[\S23]{AbramowitzStegun68}. All the coefficients are positive.
\item Evaluate the real valued Taylor series coefficients in $\phi=0$  of the function $f(\phi)$ \Tom{according to rule:} $f_{2k}=0$, $f_1=\sqrt{2d_2}$, $f_3=d_4/f_1$, and
$$
f_{2k+1} = \frac12\left(\frac{d_{2k+2}}{d_2} f_1 -\frac1k  \sum _{m=1}^{k-1} \frac{d_{2(k-m)+2}}{d_2} f_{2m+1} \right)\;.
\e{HD66}
$$
\item Evaluate the real valued Taylor series coefficients in $u=0$ of the function $\phi(u)$ \Tom{according to rule:} $\phi_{2k}=0$, $\phi_1=1/f_1$, and
$$
\phi_{2k+1} = - \sum_{m=0}^{k-1} \phi_{2m+1} f_1^{2(m-k)} F_{k-m,2m+1}
\e{HD68}
$$
where $F_{0,n}=1$, and
$$
F_{m,n} =  \sum_{k=1}^m \frac{(kn-m+k)}m \frac{f_{2k+1}}{f_1} F_{m-k,n}\;.
\e{HD70}
$$
\item Evaluate the complex valued Taylor series coefficients in $\phi=0$  of the function $\phi \,g(\phi)$ with
$$
\eqalign{
& g(\phi)  = \frac1{-\alpha'(p_\alpha(\phi))} \frac1{1-e^{q-p_\alpha(\phi)}}\cr
& = \frac{e^{i2\phi}-1}{ e^{i2\lambda\phi}\!-\!\lambda e^{i2\phi} -(1\!-\!\lambda)} 
	\cdot\frac{e^{i2\lambda\phi}-e^{i2\phi}}{(e^q\!+\!1)e^{i2\lambda\phi}-e^{i2\phi}-e^q}\;,
}
\e{HD45}
$$
\Tom{according to rule:} $ g_0=\xi_2/\nu_3$, and
$$
g_k =  \frac1{\nu_3} \left(\xi_{k+2} - \sum_{m=1}^k \nu_{m+3} g_{k-m} \right)
\e{HD72}
$$
where we used notation
$$
\eqalign{
\xi_k & = \frac{(2i)^k}{k!}\Big[(\lambda+1)^k - \lambda^k - 2^k + 1 \Big]\cr
\nu_k & = \frac{(2i)^k}{k!}\Big[ (2^k -2)(\lambda+ (1+e^q)\lambda^k)\cr
& \qquad\qquad - (1+(1+e^q)\lambda)((\lambda+1)^k-1-\lambda^k) \Big]\;.\cr
}
\e{HD74}
$$
The even coefficients are imaginary valued, and the odd coefficients are real valued. Incidentally note that, in \e{HD74}, it is $1+e^q=2$ for \ac{FA} and $1+e^q=1/P\sub{bit}$ for \ac{MD}.
\item Evaluate the Taylor series coefficients in $u=0$ of the function $c_x(u)=2u g(\phi(u))$ \Tom{according to rule:} $c_0=2g_0/\phi_1$ and
$$
c_{2k} = c_0P_k +2 \sum_{m=0}^{k-1} g_{2(k-m)} C_{m,2(k-m)-1} \phi_1^{2(k-m)-1}
\e{HD76}
$$
where $P_0=1$, $C_{0,n}=1$, and
$$
\eqalign{
P_m & = -\sum_{k=1}^m \frac{\phi_{2k+1}}{\phi_1} P_{m-k}\cr
C_{m,n} & = \sum_{k=1}^m \frac{(kn-m+k)}{m} \frac{\phi_{2k+1}}{\phi_1} C_{m-k,n}\;.
}
\e{HD78}
$$
The even coefficients are imaginary valued, and the odd coefficients are real valued.\hfill~$\Box$
\end{enumerate}
\end{procedura}
\begin{IEEEproof}
The proof is left to the reader.
\end{IEEEproof}

\subsection{Asymptotic approximations}

With a little effort, the first coefficients can be evaluated in the closed form, to obtain $c_0= ig_0(\lambda,q)$ and $c_2=-c_0 g_2(\lambda,q)$ where we used
$$
\eqalign{
g_0(\lambda,q) & =  \sqrt{\frac{(1-\lambda)}{\lambda}}\frac1{1-\lambda(e^q+1)}\cr
g_2(\lambda,q) 
 & = \frac{(1-\lambda+\lambda^2)}{12\lambda(1-\lambda)} +  \frac{\lambda e^q(1+e^q)}{(1-\lambda(1+e^q))^2} \;.
}
\e{HD80}
$$
Under the further assumption that $s_\beta < 0 < s_\alpha$, which corresponds to the request $P\sub{bit}<\lambda<\frac12$, the asymptotic approximation derived by truncating the series to the first contributions provides a result equivalent to \e{HG3} and \e{HG5}. We have

\begin{corollary}\label{hdTEO5b}
In a \ac{BSC} channel scenario where $\lambda\in(P\sub{bit},\frac12)$, the \ac{FA} and \ac{MD} probabilities \e{HD20} allow the $O(n^{-2})$ asymptotic approximations
$$
{\eqalign{
\frac1n \ln P\sub{FA}^{(1)} &= [h(\lambda) -1] \ln(2) - \frac1{2n}\ln\left(\frac{2\pi n}{ g_0^2(\lambda,0)}\right) \cr
\frac1n\ln P\sub{MD}^{(1)}  & = h(\lambda) \ln(2) + \lambda\ln(P\sub{bit}) + (1-\lambda)\ln(1-P\sub{bit}) \cr 
& \hspace*{27mm} - \frac1{2n}\ln\left(\frac{2\pi n}{ g_0^2(\lambda,\delta_0)}\right)\;,
}}
\e{HG3b}
$$
where $\lambda=\frac dn$. Furthermore, the $O(n^{-3})$ approximations
$$
{\eqalign{
\frac1n \ln P\sub{FA}^{(2)} &=\frac1n \ln P\sub{FA}^{(1)} 
	+ \frac1n\ln\Big(1- \frac1ng_2(\lambda,0)\Big)  \cr
\frac1n\ln P\sub{MD}^{(2)}  & =  \frac1n\ln P\sub{MD}^{(1)} 
	+ \frac1n\ln\Big(1- \frac1ng_2(\lambda,\delta_0)\Big)  \;,
}}
\e{HG5b}
$$
where we used \e{HD80}, also apply.\hfill~$\Box$
\end{corollary}
\begin{IEEEproof}
The result is an application of \Tom{Procedure~\ref{hdTEO5}}.
\end{IEEEproof}

One important characteristic of asymptotic approximations \e{HG3b} and \e{HG5b} is that they provide lower and upper bounds to the true \ac{FA} and \ac{MD} probabilities, so that a control on the approximation error  is available. This can be proved by mimicking Theorem~\ref{theoN8}, but the result is much more strong in the \ac{BSC} case, since the sufficient condition (equivalent to \e{QW4}) is always met and does not need to be verified.

\begin{theorem}\label{hdTEO5c}
In a \ac{BSC} channel scenario where $\lambda\in(P\sub{bit},\frac12)$ the asymptotic approximations  \e{HG3b} and \e{HG5b} of the \ac{FA} and \ac{MD} probabilities satisfy relation \e{QW2} which ensures the validity of
$$
\eqalign{
\overline{R}^{(1)} & \le \overline{R} \le \overline{R}^{(2)} \cr
\underline{P}_e^{(2)} & \le \underline{P}_e \le \underline{P}_e^{(1)} \;,
}
\e{QW6b}
$$ 
where the superscript $^{(K)}$ identifies a bound derived by use of either the $K=1$ or the $K=2$ term approximation.\hfill~$\Box$
\end{theorem}
\begin{IEEEproof}
See the Appendix.
\end{IEEEproof}

The asymptotic expressions of Theorem~\ref{hdTEO5b} also allows capturing the normal approximation, which provides a proof alternative to that of \cite[Theorem~52]{Polyanskiy10}.

\begin{theorem}\label{theo8c}
In a \ac{BSC} channel scenario, the \ac{PPV} \Tom{meta-converse}  bound is consistent with the normal approximation \e{UY2} that uses the channel dispersion coefficient
$$
V = P\sub{bit}(1-P\sub{bit}) \ln^2\left(\frac{1-P\sub{bit}}{P\sub{bit}}\right)\;,
\e{CDC2}
$$
that is, $\overline{R}=R\sub{NA} + O(1/n)$ for $n\rightarrow\infty$.\hfill~$\Box$
\end{theorem}
\begin{IEEEproof}
See the Appendix.
\end{IEEEproof}

\subsection{\ac{RCU} achievability bound}

To conclude our investigation, we finally discuss the \ac{RCU} bound which, in the \ac{BSC} scenario, can be approximated by using the tools introduced so far, which some modifications in order to be able to deal with a discrete case. 
\Tom{\bFig[t]{FI8}} 

The bound is written in the following form, which is equivalent to \cite[Theorem~33]{Polyanskiy10}.

\begin{theorem}\label{theo800}
In a \ac{BSC} channel scenario, the \ac{RCU} converse bound can be expressed as
$$
\eqalign{
\overline{P}_e & = \sum _{d=0}^{d_0}  2^{nR} P\sub{FA}(d)\, q_n(d;P\sub{bit}) + P\sub{MD}(d_0) 
}
\e{OOT2}
$$
where we used the \ac{FA} and \ac{MD} probabilities \e{HD20}, and where $d_0$ is the minimum $d$ satisfying $-\frac1n\log_2(P\sub{FA}(d))\le R$.\hfill~$\Box$
\end{theorem}
\begin{IEEEproof}
See the Appendix.
\end{IEEEproof}

An asymptotic approximation to \e{OOT2} can then be obtained by exploiting Corollary~\ref{hdTEO5b} in conjunction with the linear approximation method used in Theorem~\ref{theo28}, but suitably adapted to a discrete case. The resulting approximation is available in the following form.

\begin{theorem}\label{theo801}
In a \ac{BSC} channel scenario, for sufficiently large $n$ the \ac{RCU} bound can be approximated as
$$
\frac1n\ln \overline{P}_e  = w_n^{(0)}(\lambda) - \frac1n\ln(w_n^{(1)}(\lambda)) +O(n^{-2})
\e{WA30}
$$
where $\lambda$ is defined through the equivalence $w_n^{(2)}(\lambda)=R\ln(2)$, and where
$$
\eqalign{
w_n^{(0)}(\lambda) & = h(\lambda)\ln(2)+ \ln(1\!-\!P\sub{bit})-\lambda\delta_0 - \frac{\ln(2\pi n)}{2n} \cr
w_n^{(1)}(\lambda)
 & = \frac{(\lambda-P\sub{bit})(P\sub{bit}(1-\lambda)^2-(1-P\sub{bit})\lambda^2)}{P\sub{bit}(1-P\sub{bit})(1-2\lambda)\sqrt{\lambda(1-\lambda)}}\cr
w_n^{(2)}(\lambda) & =  [1-h(\lambda)]\ln(2)  +\frac1{2n}\ln\left(\frac{2\pi n\lambda(1-2\lambda)^2}{1-\lambda}\right) \cr
}
\e{WA32}
$$
For a correct applicability of the theorem, $w_n^{(1)}$ in \e{WA32} must be positive.\hfill~$\Box$
\end{theorem}
\begin{IEEEproof}
See the Appendix.
\end{IEEEproof}

The approximation of Theorem~\ref{theo801} finally provides a simple means to identify the consistency with the normal approximation. The following result is therefore a proof alternative to the one available in \cite[Theorem~52]{Polyanskiy10}.

\begin{theorem}\label{theo8c2}
In a \ac{BSC} channel scenario, the \ac{RCU} achievability bound is consistent with the normal approximation \e{UY2} that uses the channel dispersion coefficient \e{CDC2}, that is, $\overline{R}=R\sub{NA} + O(1/n)$ for $n\rightarrow\infty$.\hfill~$\Box$
\end{theorem}
\begin{IEEEproof}
See the Appendix.
\end{IEEEproof}

\Tom{A summary of all the above results on the \ac{BSC} channel scenario is given in compact procedural form in \fig{FI8}. Both interpretations (rate as well as error probability bound) of the \ac{RCU} bound are provided.}

\subsection{Numerical examples}

The \ac{PPV} \Tom{meta-converse} and the \ac{RCU} bounds associated with the \ac{BSC} channel are shown, together with the normal approximation, in \fig{GA8}.\bFig[t]{GA8} A \emph{hard decoding} perspective is considered where $P\sub{bit}=Q(\sqrt{\Gamma})$.  The results of Corollary~\ref{hdTEO5b} were used for the \ac{PPV} \Tom{meta-converse} bound, and, in the considered cases, Theorem~\ref{hdTEO5c} ensures validity of the plotted result. The approximation of Theorem~\ref{theo801} is instead used for the \ac{RCU} bound. Note from \fig{GA8}.(a) that, similarly to the \ac{AWGN} channel case of \fig{GA4}.(c), the \ac{PPV} \Tom{meta-converse} and the \ac{RCU} bounds are very close over a wide range of both $n$ and \ac{SNR}. A significant gap is appreciated with the normal approximation for small spectral efficiencies, and for small $n$ ($n=200$ in figure) the normal approximation is seen to be less reliable than in the \ac{AWGN} case. 

The packet error rate perspective is given in \fig{GA8}.(b) for rate $R=\frac12$, which confirms the close adherence of the \ac{RCU} and the \ac{PPV} \Tom{meta-converse} bounds over a large packet error rate range seen in the \ac{AWGN} case of \fig{GA4}.(d). Observe that the difference between the \ac{RCU} bound and the normal approximation is not strikingly significant at rate $\frac12$ (it would be much more significant at lower rates, as evident from \fig{GA8}.(a)), but \fig{GA8}.(b) completes the overview on rate $R=\frac12$ codes previously initiated in the \ac{AWGN} case of \fig{GA4}.(d), and in the BI-\ac{AWGN} case of \fig{GA6}.(b).

\section{Conclusions}

In this paper we discussed the application of the asymptotic uniform expansion approach of Temme \cite{Temme93} for the evaluation of converse and achievability bounds in the finite blocklength regime. The preliminary results available in  \cite{Erseghe15} for the \ac{AWGN} channel were generalized to a number of channels of practical interest, namely the parallel \ac{AWGN} case, the BI-\ac{AWGN} channel, and the \ac{BSC} channel, but can be in principle adapted to any memoryless channel formulation, either continuos, discrete, or mixed. The method we are using is particularly suited for evaluating the \ac{PPV} \Tom{meta-converse}  bound, for which a neat integral expression as well as a simple asymptotic series expansion is always  available. Calculation of the \ac{RCU} achievability bound is also possible, as we show in the \ac{AWGN} and \ac{BSC} case, although the final result is in general weaker.

\appendix

\begin{IEEEproof}[Proof of Theorem~\ref{theo2}]
By exploiting the fact that, in the considered context, it is $\|\B a_k\|^2 =  \Tom{\frac nK} \snr_k\sigma_w^2 $, we have
$$
\Lambda(\B a, \B b)  = \fract12 +  \fract12  \Tom{\frac1K}\sum_{k=1}^K \ln(1+\snr_k) - \frac1{2n}\Lambda'(\B a, \B b) 
\e{KK18}
$$
where
$$
\Lambda'(\B a, \B b)  = \sum_{k=1}^K \frac{\snr_k}{(1+\snr_k)} \left\|\frac{\B b_k}{\sigma_k}-\frac{(1+\snr_k)}{\snr_k}\frac{\B a_k}{\sigma_k}\right\|^2\;.
\e{KK19}
$$
Hence, we can directly work on $\Lambda'$ since it is in a linear relation with $\Lambda$. We are therefore interested in a counterpart to \e{KK12} where $\Lambda$ is replaced by $\Lambda'$. With a little effort, and by exploiting the properties of a non-central chi-squared random variable \cite[\S 2.2.3]{kay:93v2} and the same method used in \cite{Erseghe15}, we obtain  \e{KK15}. Note that in \e{KK15} the threshold level $\lambda'$ to be used in connection with $\Lambda'$ has been replaced by $\lambda$ since, operatively, they have the same meaning.
\end{IEEEproof}

\begin{IEEEproof}[Proof of Theorem~\ref{theo3}]
Denote $u=\sum_{k=1}^K  u_k\,\snr_k$ and $v=\sum_{k=1}^K  v_k\snr_k/(1+\snr_k)$ with associated \acp{PDF} $f_u(a)$ and $f_v(a)$, respectively. From \cite[eq.~29.3.81]{AbramowitzStegun68} the corresponding Laplace transforms are $F_u(s) = e^{\frac n 2\alpha(s)-n\lambda s}$ and $F_v(s) = e^{\frac n 2\beta(s)-n\lambda s}$, with associated region of convergence given in \e{KK43}. The link with the \ac{FA} and \ac{MD} probabilities in Theorem~\ref{theo2} is $P\sub{FA}(\lambda)= f_u*1(n\lambda)$ and $P\sub{MD}(\lambda)= f_v*1_-(n\lambda)$, where $*$ denotes convolution, $1(t)$ is the unit step function and $1_-(t) = 1(-t)$. By using standard Laplace transform properties \cite{Oppenheim13} the above can then be written via an inverse Laplace transform as in \e{KK30}.
\end{IEEEproof}

\begin{IEEEproof}[Proof of \Tom{Procedure~\ref{theo5}}] 
The proof uses the method exploited in \cite[Theorem~6]{Erseghe15}, to which we refer for details. The important point to observe is that $p_x^{-1}(s)$ is defined by \e{DS14}. Then, the proof is obtained by nested application of Taylor series expansion composition properties (e.g., see \cite{gradshteyn1980table}). Taylor expansions for $\alpha(s)$ and $\beta(s)$ are standard. Note that $a_2\ge0$ is a consequence of \e{KK43a}, i.e., we are looking for a minimum. The Taylor series expansion of $f(s)$ in 2) is an application of the squared-root map. The $-$ sign in $f_1$ is chosen in such a way to guarantee the correct sign of our final coefficients. The fact that the coefficients $f_m$ are imaginary valued is a consequence of \e{KK43a}. The inversion map in 3) is derived by exploiting the method used in \cite{Itskov12}. The fact that the even coefficients $p_{2m}$ are real and the odd coefficients $p_{2m+1}$ are imaginary is a consequence of the fact that $f_m$ are imaginary valued, and can be proved by induction. The result in 4) is derived from a dividing series combination rule.  The fact that the even coefficients $c_{2m}$ are imaginary and the odd coefficients $c_{2m+1}$ are real is a consequence of the properties of $p_m$, and can be easily proved by induction.
\end{IEEEproof}


\begin{IEEEproof}[Proof of Theorem~\ref{theo8}]
We first investigate the behavior at $n=\infty$, in which case  the $O(n^{-2})$ approximation \e{HG3} guarantees that $\overline{R} = -\fract12 \alpha(s_\alpha)\log_2(e)$ where $\lambda$ is set by request $\beta(s_\beta)=0$. The solution is simply $\lambda=1$ and $s_\beta=0$, and in fact $\beta(0) = 0$ for any $\lambda$, and $\beta'(0)=0$ for $\lambda=1$. These choices guarantee that $s_\alpha=\frac12$, and, in turn, that $\overline{R}=C$.

We then investigate the bound for large $n$. We preliminarily observe that the $Q$ function can be written in the form $Q(x) = e^{-\frac12x^2}q(x)$ for $x>0$, where $\ln q(x)=-\ln(\sqrt{2\pi}x) + O(1/x^2)$ \cite[eq.~26.2.12]{AbramowitzStegun68}. Hence, from \e{HG3}, the \ac{MD} probability can be approximated in the form
$$
\ln P\sub{MD} = \fract n2\beta(s_\beta) + \ln q\left(\sqrt{na_2s^2_\beta}\right) +O(1/n)\;.
\e{LX2}
$$
We then Taylor expand the functions in \e{LX2} around $\lambda=1$, which is the limit value for $n\rightarrow\infty$. To achieve our aim it is advisable to write $\beta(s;\lambda)$ in the form $2\lambda s-2h(s)$ for a suitable choice of the function $h(s)$. With this notation, we have $s_\beta(\lambda)=[h']^{-1}(\lambda)$ and $a_2(\lambda) = -h''(s_\beta(\lambda))$, which imply the approximations
$$
\eqalign{
s_\beta(\lambda) & = s_\beta^* - \frac1{a_2^*} (\lambda-1) + O((\lambda-1)^2)\cr
a_2(\lambda) & = a_2^* + O(\lambda-1)\;,\cr
}
\e{KK200}
$$
where the asterisk $^*$ denotes a quantity evaluated at $\lambda=1$. Observe that $s_\beta^*=0$ and $s_\alpha^*=\fract12$, hence from \e{DG2} it is $a_2^* = 4V$ with $V$ the channel dispersion coefficient \e{KK64bis}. As a consequence we also have
$$
\eqalign{
a_2(\lambda)s_\beta^2(\lambda) & = \frac1{4V}(\lambda-1)^2 + O((\lambda-1)^3)\cr
\beta(s_\beta(\lambda);\lambda) & = 2\Big[ \lambda s_\beta(\lambda)-h(s_\beta(\lambda)) \Big]\cr
& = -\frac{1}{4V} (\lambda-1)^2 + O((\lambda-1)^3)\;,\cr
}
\e{LX3}
$$
which allows writing \e{LX2} in the form
$$
\ln P\sub{MD} = \ln Q\left( x \sqrt{1+O(\lambda-1)}\right) +O(1/n)\;,
\e{LX4}
$$
where $x=(\lambda-1)\sqrt{{n}/{(4V)}} $. Note that, for $n\rightarrow\infty$ and $\lambda\rightarrow1$ the request $P\sub{MD}=P_e$ implies from \e{LX4} that $x\rightarrow Q^{-1}(P_e)$, and therefore $O(\lambda-1)$ is equivalent to $O(1/\sqrt{n})$. By solving for $P\sub{MD}=P_e$ we then obtain
$$
\lambda = 1 + \sqrt{\frac{4V}{n}} Q^{-1}(P_e) + O(1/n)\;,
\e{LX6}
$$
where the plus sign is consistent with the request $s_\beta(\lambda)<0$. We then inspect the \ac{FA} probability which, from \e{HG3}, can be approximated in the form
$$
-\frac1n\ln P\sub{FA} = -\fract 12\alpha(s_\alpha) + \frac{\ln (n)}{2n} +O(1/n)\;,
\e{LX8}
$$
where, from \e{KK32a}, we have
$$
-\fract12\alpha(s_\alpha(\lambda);\lambda) = -\fract12\beta(s_\beta(\lambda);\lambda) + \ln(2) \, C - \fract12(\lambda-1)\;.
\e{LX10}
$$
The approximation \e{UY2} is finally obtained by evaluating \e{LX8} with the use of \e{LX10}, the second of \e{LX3}, and \e{LX6}.
\end{IEEEproof}

\begin{IEEEproof}[Proof of Theorem~\ref{theo20}]
By exploiting \e{KK18}-\e{KK19} in definition \e{DS44}, and by recalling that $K=1$, we have
$$
\eqalign{
g(\B x,\B y) & = \P{\frac{\B y^T(\B z-\B x)}{\sigma^2}\ge 0 },\;\Tom{\hbox{where } \B z\sim \C U_{\C K}}\;.
}
\e{DS6}
$$
Then, by the circular symmetry of both $\B z$ and $\B y$ we can further derive that the bound on $P_e$ given by \e{DS2} is independent of the choice of the \ac{PDF} of $\B x$, provided that $\B x\in\C K$. In the following, with no loss in generality, we therefore assume that $\B x$ is a vector with the first entry set to $\sqrt{nP}$ and the rest to $0$. In \e{DS6} we then exploit the fact that $\B z$ can be written in the form
$$
\B z= \frac{\B g}{\|\B g\|}\sqrt{nP}\;,\quad \B g\sim\C N(\B 0_{n},\B I_{n})\;,
\e{DS8}
$$
and that, because of the circular symmetry of $\B z$, the substitution 
$$
\frac{\B y^T\B z}{\sigma^2} \rightarrow  \frac{\|\B y\|}{\sigma} \eta \sqrt{n \snr}\;,\quad
\eta = \frac{g_{1}}{\|\B g\|}\;.
\e{DS10}
$$
provides an equivalent result. With this notation, \e{DS20} is valid by setting
$$
\tau = g_{1} \sqrt{\frac{n-1}{\|\B g\|^2-g_{1}^2}}\;, \quad \B q=\frac{\B y}{\sigma\sqrt{\snr/n}}\;,
$$
where $\tau$ is by definition a t-distributed variable, and where $\B q$ is Gaussian as in \e{DS2b}. This proves the theorem.\end{IEEEproof}

\begin{IEEEproof}[Proof of Theorem~\ref{theo26}]
We note that $\rho$ can be written in the form
$$
\rho = \frac{d}{\sqrt{n-1+d^2}}\;,\quad d = q_1\sqrt{\frac{n-1}{\|\B q\|^2-q_1^2}}
\e{AV20}
$$
where $d$ is a non central t-distributed random variable of order $n-1$ an non-centrality parameter $\sqrt{n\snr}$. Hence, with some effort, we also find that
$$
\P{\rho\le a} = \P{d\le\frac{a\sqrt{n-1}}{\sqrt{1-a^2}}}
$$
Then, by exploring the \ac{PDF} of a t distributed random variable of order $\nu$ and non-centrality parameter $\mu$, namely (see \cite[p.~177]{scharf1991statistical} and \cite[eq.~19.5.3]{AbramowitzStegun68}, and be advised that expression \cite[eq.~26.7.9]{AbramowitzStegun68} is not correct)
$$
\eqalign{
f_t(a) & = \frac{2\Gamma(\nu+1)}{\sqrt{\nu \pi} \Gamma(\frac\nu2)} U\left(\nu+\fract12;-\frac{\mu a}{\sqrt{\nu+a^2}}\right)
	 \left(\frac{\nu/2}{\nu+a^2}\right)^{\frac{\nu+1}2} \cr
 & \hspace*{20mm}\cdot  e^{-\frac14\mu^2\left(1+\frac\nu{\nu+a^2}\right)}
 }
\e{AV21}
$$
where $U(a;x)$ is Weber's form for the parabolic cylinder function (decreasing to $0$ for large $x$), we obtain \e{AV22}. With a little effort, from  \cite[eq.~19.10 and 6.1.40]{AbramowitzStegun68} we have 
$$
\eqalign{
u_n(a) & =  \fract12\big(1-\fract3{n}\big)\ln(1-a^2)+(\alpha a)^2-2\alpha^2 \cr
 & \qquad -\fract12\ln\big(1-\fract1n\big) +\alpha a\sqrt{1+(\alpha a)^2-\fract1{2n}}\cr
 & \qquad  + \big(1-\fract1{2n}\big)\sinh^{-1}\Big(\alpha a\Big/\sqrt{1-\fract1{2n}}\Big)  +\fract1{2n}\ln(n)\cr
 & \qquad  -\fract1{2n}\ln\big(2\pi e\sqrt{1+(\alpha a)^2}\big) + O(n^{-2})\;.
}
\e{AV30}
$$
Equation \e{AV302} is then obtained as a rearrangement of \e{AV30}, by neglecting the $O(n^{-2})$ contributions.
\end{IEEEproof}

\begin{IEEEproof}[Proof of Theorem~\ref{theo22}]
The statistical properties of $\eta$ can be identified in the closed form. From \e{DS20} the \ac{CDF} of $\eta$ can be written in the form
$$
\P{\eta\le a} = \P{\tau\le \frac{a\sqrt{n-1}}{\sqrt{1-a^2}}}
\e{DD16}
$$
for $0\le a<1$, while for $-1<a<0$ it simply is $\P{\eta\le a}=1-\P{\eta\le -a}$ because of the symmetry of $\eta$. Hence the \ac{PDF} of $\eta$ is (see \cite[eq.~26.7.1]{AbramowitzStegun68})
$$
f_{\eta}(a) = \frac{\Gamma(\fract{n}2)}{\Gamma(\fract12)\Gamma(\fract{n-1}2)} (1-a^2)^{\frac{n-3}2} \;,
\e{DD18}
$$
with $a\in(-1,1)$, and the corresponding Laplace transform is (see \cite[eq.~9.6.18 and 9.6.47]{AbramowitzStegun68}).
$$
\eqalign{
F_{\eta}(s)
	& =\rule{0mm}{1em}_0F_1(;\fract{n}2;\fract14s^2)\cr
	 & = \exp\left(\fract{n}2 G_{\frac{n}2}(2s/n)\right)\;.
}
\e{DD20}
$$
The result of \e{DS22} is then a counterpart to the derivation of the \ac{MD} probability in Theorem~\ref{theo3}.
\end{IEEEproof}

\begin{IEEEproof}[Proof of Theorem~\ref{theoq2}]
From \cite[eq.~9.6.26, 9.6.47, 9.7.7, and 9.7.9]{AbramowitzStegun68} we have that the function $G_\nu$ can be written in the form
$$
\eqalign{
& G_\nu(s) \cr
& =\sqrt{1+s^2} +  \fract1\nu\ln\left(\frac{\Gamma(\nu)}{\nu^{\nu-\frac12}\sqrt{2\pi}}\right) - \fract1{2\nu}\ln(\sqrt{1+s^2}) \cr
& \quad  -\Big(1-\fract1\nu\Big)\ln\Big(\frac{1+\sqrt{1+s^2}}2\Big) + \fract1\nu\ln\left(1 + \sum_{k=1}^\infty g_k(t)\nu^{-k}\right)
}
\e{CR6}
$$
where $t=1/\sqrt{1+s^2}$, and where $g_k(t)= (v_k(t)+t\,u_k(t))/(1+t)$ with $v_k(t)$ and $u_k(t)$ the polynomials defined in \cite[eq.~9.3.9-14]{AbramowitzStegun68}. By construction, $g_k(t)$ is a polynomial of order $k$ in $t$, and for the first orders we have
$$
\eqalign{
g_1(t) & = \frac{-5t^3+12t^2-9t}{24}\cr
g_2(t) & = \frac{385t^6  -840 t^5+  378t^4+   216t^3  -135t^2}{1152}\;.
}
\e{CR8}
$$
By limiting the serie to the term $k=0$, and by exploiting \cite[eq.~6.1.40]{AbramowitzStegun68}, we obtain the asymptotic expression \e{CR10}.
\end{IEEEproof}

\begin{IEEEproof}[Proof of Theorem~\ref{theo28}]
Recall that $\lambda$ is defined by $v_n(\lambda)=R\ln(2)$. The theorem is proved by assuming that the function $u_n(a)$ is increasing for $a\in[-1,\lambda]$, and that the function $u_n(a)-v_n(a)$ is decreasing for $a\in[\lambda,1]$, which (because of the shape of the involved functions) correspond to assuming that the function maxima lie outside the integration interval, which also implies a small error probability $P_e<\frac12$. Incidentally note that, from the $O(n^{-1})$ approximations that can be derived from \e{AV302} and \e{AV10b}, the function maxima can be approximated in the form
$$
\eqalign{
\argmax_a u_n(a) & = \sqrt{\frac\snr{1+\snr}}+ O(n^{-1})\cr
\argmax_a u_n(a)-v_n(a) & = \sqrt{\frac{\sqrt{4+\snr^2}-2}{\snr}}+ O(n^{-1})\;.
}
\e{AV40}
$$
Since the Shannon bound guarantees $R<\frac12\ln(1+\snr)$, from \e{AV14bi} we have 
$$
\lambda<\sqrt{\frac\snr{1+\snr}}+ O(n^{-1})
\e{AV41}
$$
and the assumption is practically guaranteed for the first contribution in \e{AV32}. The assumption $v_n'(\lambda) > u_n'(\lambda)$ further guarantees that it is satisfied also for the second contribution in \e{AV32}.

Given the above, for the first contribution in \e{AV32} we substitute variable $a\in[-1,\lambda]$ with variable $x\in(-\infty,0]$ such that $u_n(a)=u_n(\lambda)-x^2$, that is $x = f(a)=-\sqrt{u_n(\lambda)-u_n(a)}$. This provides the equivalent integral expression
$$
\int_{-1}^\lambda e^{n u_n(a)} da = \int_{-\infty}^0 e^{n u_n(\lambda)-n x^2} \underbrace{\frac{-2x}{u_n'(f^{-1}(x))}}_{h(x)} dx\;.
\e{AV42}
$$
By Taylor expansion of the function $h(x)$ in $x=0$, and by use of \cite[eq.~7.4.4-5]{AbramowitzStegun68}, we obtain
$$
\int_{-1}^\lambda e^{n u_n(a)} da = e^{n u_n(\lambda)}  \sum_{k=0}^\infty h_k \frac{(-1)^k\Gamma(\frac{k+1}2)}{2n^{\frac{k+1}2}}\;.
\e{AV44}
$$
By then observing that $h_0=0$, $h_1=-2/u_n'(\lambda)$, and $h_2=0$, the approximation
$$
\int_{-1}^\lambda e^{n u_n(a)} da = e^{n u_n(\lambda)}\left(\frac{1}{n u_n'(\lambda)} + O(n^{-2})\right)
\e{AV46}
$$ 
holds. We can proceed equivalently for the second integral in \e{AV32}, where we define $g_n(a)=u_n(a)-v_n(a)+R\ln(2)$ for compactness. In this case the integration variable is $x\in[0,\infty)$ in the relation $x=f(a)=\sqrt{g_n(\lambda)-g_n(a)}$ with $a$. This ensures
$$
\int_\lambda^1 e^{n g_n(a)} da =  \int_0^\infty e^{n g_n(\lambda)-n x^2} \underbrace{\frac{-2x}{g_n'(f^{-1}(x))}}_{h(x)} dx
\e{AV48}
$$
where the first Taylor coefficients are $h_0=0$, $h_1=-2/g_n'(\lambda)$, and $h_2=0$. By use of \cite[eq.~7.4.4-5]{AbramowitzStegun68} we then obtain
$$
\int_\lambda^1 e^{n g_n(a)} da = e^{n g_n(\lambda)} \left(\frac{1}{n (-g_n'(\lambda))} + O(n^{-2})\right)
\e{AV50}
$$
where $g_n(\lambda)=u_n(\lambda)$ because of the definition of $\lambda$, and where $-g_n'(\lambda)= v_n'(\lambda)-u_n'(\lambda)>0$ by assumption. Then
$$
\eqalign{
\frac1n\ln \overline{P}_e & = u_n(\lambda) - \frac{\ln(n)}n  \cr
& \qquad +\frac1n\ln\left(\frac{v_n'(\lambda)}{u_n'(\lambda) [v_n'(\lambda)-u_n'(\lambda)]}\right) +O(n^{-2})
}
\e{AV33bi}
$$ 
follows from the sum of \e{AV46} and \e{AV50}.  The $O(n^{-2})$ approximation \e{AV302} should be used for $u_n$, while for derivatives the $O(n^{-1})$ approximations (derived from \e{AV302} and \e{CR12}) 
$$
\eqalign{
v_n'(a) & = \frac a{1-a^2}+ O(n^{-1})\cr
u_n'(a) & =  \frac {-a}{1-a^2} + 2\alpha\Big(\alpha a + \sqrt{1+(\alpha a)^2}\Big)+ O(n^{-1}) \cr
}
\e{AV33b}
$$
can be used. Equation \e{AV33} is a compact rewrite of \e{AV33bi} using \e{AV33b} and \e{AV302}-\e{AV304}. The assumption  $v_n'(\lambda) > u_n'(\lambda)$ corresponds to $w^{(0)}(\lambda)>0$.\end{IEEEproof}

\begin{IEEEproof}[Proof of Theorem~\ref{theo28b}]
The proof mimics the one of Theorem~\ref{theo8}. We first investigate the behavior at $n=\infty$, in which case \e{AV10b} and \e{AV33} guarantee that $\underline{R} = -\fract12 \log_2(1-\lambda^2)$ where $\lambda$ is set by request $u^{(0)}(\lambda)=0$. It can be verified that the result is $\lambda^*=\sqrt{\snr/(1+\snr)}$ and $\underline{R}=C$. We then investigate the Taylor series expansion of \e{AV33} at $\lambda=\lambda^*$. By standard methods we obtain
$$
\eqalign{
u^{(0)}(\lambda) & = -\frac1W(\lambda-\lambda^*)^2 + O((\lambda-\lambda^*)^3)\cr
w^{(0)}(\lambda) & = -\frac1{\sqrt{W}}(\lambda-\lambda^*) + O((\lambda-\lambda^*)^2)\cr
}
$$
where $W=(2+\snr)/(2(1+\snr)^3)=V/(\snr(1+\snr))$. By squaring the approximation on $w^{(0)}(\lambda)$, the result we obtain is equivalent to \e{KK200}. Hence, by exploiting the same method that was used in Theorem~\ref{theo8} we can say that (compare with \e{LX6})
$$
\lambda = \lambda^* - \sqrt{\frac{W}{n}} Q^{-1}(P_e) + O(n^{-1})\;,
\e{LX6bbb}
$$
where the minus sign is consistent with the request $w^{(0)}(\lambda)>0$. The normal approximation then follows by observing that, from \e{AV10b}, it is
$$
\eqalign{
v_n(\lambda) & = C\ln(2) + \sqrt{\snr(1+\snr)} (\lambda-\lambda^*) + \frac{\ln(n)}{2n} \cr
 & \qquad +  O((\lambda-\lambda^*)^2) +  O(n^{-1})\;,
}
\e{CHE2}
$$
and by substitution of \e{LX6bbb}.
\end{IEEEproof}

\begin{IEEEproof}[Proof of Theorem~\ref{sdTEO1}]
We preliminarily observe that, by substitution of \e{SD4} and \e{SD8} in \e{KK10}, we obtain
$$
\Lambda(\B a, \B b) = \ln2 - \frac1n\sum_{i=1}^n h(\snr a_i b_i)
\e{SD10}
$$
where $h$ is defined in the enunciation of the theorem. Similarly to the proof of Theorem~\ref{theo2}, we can therefore linearly modify $\Lambda$, and introduce an alternative threshold $\lambda$ defined through the linear relation $\lambda_R = \ln2 - \lambda$. By use in \e{KK12}, we obtain
$$
\eqalign{
P\sub{FA}(\B a) & = \P{\sum_{i=1}^n h(\snr a_i y_i)\le n \lambda}\!,\;\Tom{\hbox{where }\B y\sim p_y}\cr
P\sub{MD}(\B a) & = \P{\sum_{i=1}^n h(\snr a_i y_i)> n \lambda}\!,\Tom{\hbox{where } \B y\sim p_{y|x},\B x=\B a},
}
\e{SD10bis}
$$
from which \e{SD20} straightforwardly follows by inspecting the statistical properties of products $\snr a_iy_i$.
\end{IEEEproof}

\begin{IEEEproof}[Proof of Theorem~\ref{theo8bi}]
We preliminarily prove that the \ac{PPV} \Tom{meta-converse} bound provides capacity at $n=\infty$. To this end, we observe that the following equivalences $H(0)=1$ and $H'(0)=(C-1)/\log_2(e)$ hold. Therefore, from Theorem~\ref{theo1} and \e{HG3} with $n=\infty$, we have that $\lambda$ and $s_\beta$ are set by, $\beta(s_\beta)=0$ and $\beta'(s_\beta)=0$, which provides $s_\beta=0$ and $\lambda=(1-C)/\log_2(e)$. Hence, it also is $s_\alpha=1$. Therefore the bound becomes $\overline{R}=-\frac12\alpha(s_\alpha)\log_2(e)$ which provides $\overline{R}=C$, and the result is proved.

The rest of the proof mimics the one of Theorem~\ref{theo8}, by using the notation valid in the binary codewords case, and by observing that the function $h$ used in the proof of Theorem~\ref{theo8} corresponds to the function $-\ln(H(s))$. Equation \e{LX2} is still valid because of Theorem~\ref{sdTEO2}, and so is \e{KK200} but we should consider that the limit value for $\lambda$ is $\lambda^*=(1-C)/\log_2(e)$. Hence \e{KK200} rewrites as
$$
\eqalign{
s_\beta(\lambda) & = s_\beta^* - \frac1{a_2^*} (\lambda-\lambda^*) + O((\lambda-\lambda^*)^2)\cr
a_2(\lambda) & = a_2^* + O(\lambda-\lambda^*)\;,\cr
}
\e{KK200b}
$$
where the asterisk $^*$ denotes a quantity evaluated at $\lambda^*$, e.g., $s_\beta^*=0$, $s_\alpha^*=1$. Note also, that, differently from the proof of Theorem~\ref{theo8}, it is $a_2^* = \frac12\beta''(0)=V$, so that expressions hold provided that the following mappings are applied, namely: $4V\rightarrow V$, and $\lambda-1\rightarrow\lambda-\lambda^*$. Hence, \e{LX3} turns into
$$
\eqalign{
a_2(\lambda)s_\beta^2(\lambda) & = \frac1{V}(\lambda-\lambda^*)^2 + O((\lambda-\lambda^*)^3)\cr
\beta(s_\beta(\lambda);\lambda) & =  -\frac{1}{V} (\lambda-\lambda^*)^2 + O((\lambda-\lambda^*)^3)\;,\cr
}
\e{LX3bis}
$$
and the counterpart to \e{LX6} becomes
$$
\lambda = \lambda^* + \sqrt{\frac{V}{n}} Q^{-1}(P_e) + O(1/n)\;.
\e{LX6bis}
$$
In the binary codewords context \e{LX10} is then reinterpreted in the form
$$
-\fract12\alpha(s_\alpha(\lambda);\lambda) = -\fract12\beta(s_\beta(\lambda);\lambda) + \ln(2) -\lambda \;,
\e{LX10bis}
$$
so that by substitution in \e{LX8} we obtain  \e{UY2}.
\end{IEEEproof}

\begin{IEEEproof}[Proof of Theorem~\ref{theo40pre}]
From \e{SD10} we have
$$
g(\B x,\B y) = \P{ \sum_{i=1}^n h(\snr x_i y_i) \ge \sum_{i=1}^n h(\snr z_i y_i)
	}
\e{SDD22}
$$
\Tom{where $\B z\sim \C U_{\C K}$}. We then consider that $y_i$ can be written in the form $y_i=x_iw_i/\snr$ where $w_i\sim\C N(\snr,\snr)$, which provides
$$
\eqalign{
g(\B w)  = \P{ \sum_{i=1}^n h( w_i) \ge \sum_{i=1}^n h( z_i w_i)
	}\!,\;\Tom{\hbox{where } \B z\sim \C U_{\C K}}
}
\e{SDD23}
$$
since $x_i^2=1$ and since $x_iz_i$ has the same statistical description of $z_i$. By further exploiting $h(x)-h(-x)=-2x$, we obtain \e{SDD20} and \e{SDD21}.
\end{IEEEproof}

\begin{IEEEproof}[Proof of Theorem~\ref{theo40}]
Denote $g(\B a)=\int_{R^c} p_{y|x}(\B b|\B a) d\B b$, and observe that 
$$
\int _{\C R} p_y(\B b) d\B b=  \sum_{\Bs a\in\C K} \frac1{2^n} \int _{\C R} p_{y|x}(\B b|\B a) d\B b = 1 -  \sum_{\Bs a\in\C K} \frac{g(\B a)}{2^n}\;.
\e{KB10}
$$
Since $g(\B a)\le1-\tau$, it also is $\kappa(\tau)\ge\tau$. The lower bound can be reached by choosing a circularly symmetric region such that $g(\B a)=1-\tau$, for all $a\in\C K$. Then \e{KB12} follows by exploiting the equality $\kappa(\tau)=\tau$ in \e{KB2}.
\end{IEEEproof}

\begin{IEEEproof}[Proof of Theorem~\ref{hdTEO1}]
We preliminarily observe that equations \e{ADO2}-\e{ADO2bb} provide the correct interpretation of the Neyman-Pearson criterion when dealing with discrete  random variables. We then explicit the log-likelihood function \e{KK10} under \e{HD4}, to have
$$
\Lambda(\B a, \B b) = \ln(2(1-P\sub{bit})) - \frac1n\|\B b-\B a\|_H \ln\left(\frac{1-P\sub{bit}}{P\sub{bit}}\right)
\e{HD14}
$$
where the latter logarithm is guaranteed to be positive since $P\sub{bit}<\frac12$. Hence, the \ac{FA} and \ac{MD} probabilities \e{KK12} can be written in the form
$$
\eqalign{
P\sub{FA}(\B a,\lambda) & = \P{\|\B y-\B a\|_H\le n\lambda}\!,\;\Tom{\hbox{where } \B y\sim p_y}\cr
P\sub{MD}(\B a,\lambda) & = \P{\|\B y-\B a\|_H> n\lambda}\!,\;\Tom{\hbox{where } \B y\sim p_{y|x}, \B x=\B a}\;.
}
\e{HD16}
$$
We then observe that $\|\B y-\B a\|_H=\sum_i \|y_i-a_i\|_H$ holds in \e{HD16}, with statistically independent $y_i$'s. By making explicit the statistical description of the binary variables $\|y_i-a_i\|_H$ in \e{HD16}, and by observing that these are independent of the value of $\B a$, we then have \e{HD20} where we set $\lambda=\frac dn$. 
\end{IEEEproof}

\begin{IEEEproof}[Proof of Theorem~\ref{hdTEO2}]
The proof can be carried out as already seen in Theorem~\ref{theo3}, by noting that the Laplace transforms of \acp{PDF} of $u_i$'s and $v_i's$ are, respectively, of the form  $\fract12 + \fract12 e^{-s}$, and $1-P\sub{bit} +P\sub{bit} e^{-s}$. This provides $\alpha$ as in \e{HD32}, and $\beta(s) = 2\lambda s+ 2\ln(1+e^{-(s+\delta_0)})+2\ln(1-P\sub{bit})$, which is equivalent to the second of \e{HD32}. Note that application of the Laplace transform method provides probabilities \e{HD20} only for non integer values of $d$. In fact, at points where the function is discontinuous, i.e., for integer values of $d$,  the inverse Laplace transform equals the average of the left and right limits.
\end{IEEEproof}

\begin{IEEEproof}[Proof of Theorem~\ref{hdTEO2bis}]
In order to identify the \ac{FA} probability in \e{HD20} we need to investigate the right limit $d^+$ for integer $d$, which corresponds to setting $\lambda = \frac dn+\epsilon$, with $\epsilon>0$ as small as desired, in Theorem~\ref{hdTEO2}. By now exploiting \e{HD37c}, and the equivalence $\alpha(s+i2k\pi)=\alpha(s)+ 2\lambda\cdot i2k\pi$, then the \ac{FA} probability can be written in the form
$$
\eqalign{
P\sub{FA}(d) 
 & = 1(-s_\alpha) + \frac1{i2\pi} \sum_{k=-\infty}^{+\infty} \int_{\C A_\alpha} e^{\frac n2\alpha(s)}\frac{e^{n\lambda\cdot i2k\pi}}{s+ i2k\pi} \;ds\cr
 & = 1(-s_\alpha) + \frac1{i2\pi} \int_{\C A_\alpha} e^{\frac n2\alpha(s)} \left(\sum_{k=-\infty}^{+\infty} \frac{e^{n\lambda\cdot i2k\pi}}{s+ i2k\pi}\right) \;ds
}
\e{UH2}
$$
where the second equivalence was obtained by exchanging integration and sum. The function in brackets in \e{UH2} can be evaluated in the closed form by means of standard Fourier transform/series properties. The rationale is the following. Start from the Fourier transform pairs
$$
\eqalign{
1(t) e^{qt} & \quad\longrightarrow\quad \frac1{i2\pi f - q}\quad \Re(q)<0\cr
-1(-t) e^{qt} & \quad\longrightarrow\quad \frac1{i2\pi f - q}\quad \Re(q)>0\;.
}
\e{HD50}
$$
Recall that a sampling in frequency with sampling period $F=1$ corresponds to a periodic repetition in time with period $T_p=1/F=1$, with corresponding inverse Fourier transform relations
$$
\sum_{k=-\infty}^\infty  \frac{e^{i2\pi kt}}{i2\pi k-q} = \cases{
\displaystyle\sum_{k=-\infty}^\infty 1(t+k) e^{q(t+k)} &, $\Re(q)<0$\cr
\displaystyle-\sum_{k=-\infty}^\infty 1(-t+k) e^{q(t-k)}\!\!\!&, $\Re(q)>0\,$.\cr
}
\e{HD52}
$$
In both cases the result is
$$
\sum_{k=-\infty}^\infty  \frac{e^{i2\pi kt}}{i2\pi k-q} =
\cases{ 
\displaystyle\frac{1 + e^{q}}{2(1-e^q)} &, $t$ an integer\cr
\rule{0mm}{7mm}\displaystyle\frac{e^{(t-\lfloor t\rfloor)\,q}}{1-e^q} &, $t$ not an integer \cr
}
\e{HD54}
$$
for $\Re(q)\neq 0$, but the result is valid also for $\Re(q)=0$ by virtue of continuity. We observe that, for a sufficiently small $\epsilon$, the application of \e{HD54} to \e{UH2} provides a contribution of the form  $1/(1-e^q)$ since  $t=n\lambda=d+n\epsilon$ and $t-\lfloor t\rfloor = n\epsilon\simeq 0$. We therefore obtain
$$
\eqalign{
P\sub{FA}(d) & = 1(-s_\alpha)+ \frac1{i2\pi} \int_{\C A_\alpha} \frac{e^{\frac n2\alpha(s)} }{1-e^{-s}} \;ds\cr
P\sub{MD}(d) & = 1(s_\beta)-\frac1{i2\pi} \int_{\C A_\beta} \frac{e^{\frac n2\beta(s)} }{1-e^{-s}} \;ds\cr
}
\e{UH2b}
$$
the result for the \ac{MD} probability being derived in a perfectly equivalent way. Then, \e{HDW2} is obtained by explicitly using the path expressions.
\end{IEEEproof}

\begin{IEEEproof}[Proof of Theorem~\ref{hdTEO5c}]
The theorem proof mimics the one of Theorem~\ref{theoN8}, where inequality \e{QW4} implies \e{QW2}, and, in turn, \e{QW2} implies \e{QW6}. In the \ac{BSC} context, the equivalent to inequality \e{QW4} is
$$
1- g_2(\lambda,q) u^2(\phi) \le \frac1{g_0(\lambda,q)} \Im[c_x(u(\phi))] \le 1\;,
\e{HD90}
$$
where $u^2(\phi)=2d(\phi)$, and implies \e{QW2}. In turn, it can be easily verified that \e{QW2} implies \e{QW6} also under adoption of \e{ADO2}-\e{HD20}. Our aim is therefore to prove \e{HD90}.

We separate the main function in \e{HD90} in the form $\Im[c_x(u(\phi))]/g_0(\lambda,q) =f_1(\phi)f_2(\phi)$ with
$$
\eqalign{
f_1(\phi) & = \frac{2u(\phi)\,\sqrt{\lambda(1-\lambda)}}{-\alpha'(s(\phi)\!+\!i\phi)(s'(\phi)\!+\!i)}
 = \frac{\sqrt{2d(\phi)\lambda (1-\lambda)}}{d'(\phi)\,{\rm sign}(\phi)} \cr
f_2(\phi) & = \Im\left[\frac{s'(\phi)+i}{1-e^{q-s(\phi)-i\phi}}\right] \cdot\frac{(1-\lambda(e^q+1))}{(1-\lambda)}\cr
 }
\e{HD200}
$$
and $s(\phi)=\ln(v(\phi))$. With a little effort we can  write
$$
\eqalign{
f_1(\phi) & =  \frac{\sqrt{1+ D(\phi)}}{1 + C(\phi) } \cr
f_2(\phi) & =  1 - b_2 \frac{ (1-\lambda) A(\phi)+ \lambda B(\phi)}{ 1+  \alpha A(\phi)  + \beta B(\phi)}
}
\e{HD202}
$$
with
$$
b_2   =  \frac{\lambda e^q(1+e^q)}{(1-\lambda(1+e^q))^2} 
\e{HD201}
$$
and
$$
\eqalign{
\alpha & = \frac{(1+e^{q}) (1-\lambda)^ 2}{(1-\lambda(1+e^q))^2}\cr
A(\phi) & = \frac{\sinc^2((1-\lambda)\phi)}{\sinc^2(\phi)} -1\cr 
C(\phi) & = \frac{d'(\phi)}{2d_2\phi}-1\cr
 & = \sum_{k=1}^\infty \frac{d_{2k+2}}{d_2} (k+1) \phi^{2k}
}\quad\eqalign{
\beta & = \frac{e^q (1+e^q) \lambda^2}{(1-\lambda(1+e^q))^2}\cr
B(\phi) & =\frac{\sinc^2(\lambda\phi)}{\sinc^2(\phi)}  -1\cr    
D(\phi) & = \frac{d(\phi)}{d_2\phi^2}-1\cr
 & = \sum_{k=1}^\infty \frac{d_{2k+2}}{d_2} \phi^{2k}\;.
}
\e{HD206}
$$
Note that all constants and functions in \e{HD201} and \e{HD206} are positive, and in particular it is $0\le A(\phi)\le B(\phi)$ because of $\sinc$ properties, and $2D(\phi)\le C(\phi)$ since all the Taylor coefficients $d_{2k}$ are positive. Incidentally, this latter property ensures $f_1(\phi)\le 1$, while positivity of  coefficients and functions ensures $f_1(\phi)\ge0$ and $f_2(\phi)\le 1$. This proves the upper bound in \e{HD90}, since $f_1(\phi)f_2(\phi)\le f_1(\phi)\le 1$. 

For the lower bound we employ a different strategy for the \ac{FA} and \ac{MD} probabilities. Under \ac{MD}, that is with $1+e^q=P\sub{bit}^{-1}$ and $P\sub{bit}<\lambda<\frac12$, we first prove that 
$$
f_2(\phi) \ge 1 - 2 b_2 d(\phi)\;.
\e{HD205}
$$
To do so, we exploit inequalities $\alpha\ge0$ and $\beta\ge1$, which provide $\alpha A(\phi)+\beta B(\phi) \ge B(\phi)\ge  A(\phi)$. We then have
$$
\eqalign{
2 d(\phi) & = (1-\lambda)\ln(1\!+\! A(\phi)) + \lambda \ln (1\!+\!B(\phi))\cr
 & \ge (1-\lambda) \frac{A(\phi)}{1+A(\phi)}+ \lambda \frac{B(\phi)}{1+B(\phi)}\cr
  & \ge \frac{ (1-\lambda) A(\phi)+ \lambda B(\phi)}{ 1+  \alpha A(\phi)  + \beta B(\phi)}
}
\e{HD205b}
$$
where we used $\ln(1+x)\ge x/(1+x)$ in the first inequality. By substitution of \e{HD205b} in the second of \e{HD202} we obtain  \e{HD205}. We then observe that 
$$
f_1(\phi) \ge 1 - 2 a_2  d(\phi) \;,\quad a_2 =  \frac{3d_4}{4d_2^2} = \frac{(1-\lambda+\lambda^2)}{12\lambda(1-\lambda)}\;,
\e{HD300}
$$
an inequality which we verified numerically over $\phi\in(0,\pi)$ and $\lambda\in(0,\frac12)$. The lower bound in \e{HD90} is then a consequence of the validity \e{HD205} and \e{HD300}, of the property $0\le f_1(\phi)\le 1$, and of the equivalence $a_2+b_2 = g_2(\lambda,q)$. Under \ac{FA}, where $e^q=1$ and $0<\lambda<\fract12$, we numerically verified that $f_1(\phi)f_2(\phi) \ge 1 -2g_2(\lambda,q) d(\phi)$ holds for $\phi\in(0,\pi)$ and $\lambda\in(0,\frac12)$. This completes the proof.
\end{IEEEproof}

\begin{IEEEproof}[Proof of Theorem~\ref{theo8c}]
We mimic Theorem~\ref{theo8}, but the proof here is complicated by the fact that we are dealing with a discrete distribution. Let $f(x)$ be a strictly decreasing function which for integer valued $x$ assumes the values $\frac1n\ln(P\sub{MD}(x))$, $x\in\M N$. Then, the value $d$ can be obtained from value $d_0$ satisfying $f(d_0)=\frac1n\ln(P_e)$, in such a way that $d=\lceil d_0\rceil = d_0+O(1)$ and $\lambda = \frac{d_0}n + O(n^{-1})$. Another important result that will be useful is that \e{ADO2} ensures 
$$
\overline{R} = -\fract1n\log_2( P\sub{FA}(d))+O(n^{-1})\;.
\e{LX2bb}
$$
Now, for $n=\infty$ we have $f(d)=\beta(s_\beta)$ and the constraint is $f(d_0)=0$ which provides $\lambda_0=\frac{d_0}n = P\sub{bit}$ and also $\lambda=\lambda_0$. Then, from \e{HG3b} and \e{LX2bb}, and by inspection of \e{HD39}, we have $\overline{R}=-\frac12\log_2(e)\alpha(s_\alpha)=1-h(P\sub{bit})=C$ where we also exploited the fact that $n=\infty$ to remove the $O(n^{-1})$ contributions. We then inspect the case $n<\infty$. In the \ac{BSC} case the equivalent to \e{LX2} is
$$
n f(d_0) = \fract n2\beta(s_\beta) + \ln q\left(\sqrt{ng_0^{-2}(\lambda_0,\delta_0)}\right) +O(1/n)\;,
\e{LX2b}
$$
as can be derived by comparing \e{HG3} and \e{HG3b}. The values of $g_0(\lambda_0,\delta_0)$ and $\beta(s_\beta)$ are available from, respectively, \e{HD80} and \e{HD39}, and imply the equivalences
$$
\eqalign{
g_0^{-2} (\lambda_0,\delta_0) 
 & = \frac{1}{W} (\lambda_0-P\sub{bit})^2 + O((\lambda_0-P\sub{bit})^3)\cr
\beta(s_\beta(\lambda_0);\lambda_0) 
 & =  -\frac{1}{W} (\lambda_0-P\sub{bit})^2 + O((\lambda_0-P\sub{bit})^3)
}
\e{KH400}
$$
with $W = P\sub{bit}(1-P\sub{bit})$. By the same arguments that lead from \e{LX3} to \e{LX4} and \e{LX6}, we obtain
$$
\lambda_0 = P\sub{bit} + \sqrt{\frac{W}{n}} Q^{-1}(P\sub{bit}) + O(1/n)\;,
\e{LX6bi}
$$
and therefore $\lambda=\lambda_0+O(1/n)$, so that the two have the same $O(1/n)$ approximation. We then observe that the logarithmic version of the \ac{FA} probability is equivalently expressed in the form  \e{LX8}, and that \e{LX2bb} holds. Hence, the theorem is proved by observing (from \e{HD39}) that
$$
\eqalign{
-\fract12\alpha(s_\alpha(\lambda);\lambda) & = -\fract12\beta(s_\beta(\lambda);\lambda)  
	+  \ln(2) \cr & \qquad + \lambda\ln(P\sub{bit}) + (1-\lambda)\ln(1-P\sub{bit})\;,
}
\e{LX6tri}
$$
and by final substitution of the second of \e{KH400} and \e{LX6bi} in \e{LX6tri}.
\end{IEEEproof}

\begin{IEEEproof}[Proof of Theorem~\ref{theo800}]
By investigation of \e{HD14} we have
$$
\eqalign{
g(\B x,\B y) & = \P{\|\B x-\B y\|_H\ge \|\B z-\B y\|_H}\!,\;\Tom{\hbox{where } \B z\sim\C U_{\C K} }\cr
& = \P{\|\B z\|_H \le \|\B x-\B y\|_H}\!,\;\Tom{\hbox{where } \B z\sim\C U_{\C K} } \cr
& = g(\|\B x-\B y\|_H)\;,
}
\e{OOT4}
$$
where we exploited the fact that $\|\B z\|_H$ and $\|\B z-\B y\|_H$ have the same statistical description. Then, also the \ac{RCU} bound \e{DS2} is independent of $\B x$, and we can therefore choose $\B x=\B 0$, to have
$$
\eqalign{
\overline{P}_e & = \E{\min(1,2^{nR} g(\|\B y\|_H)) }\!,\;\Tom{\hbox{where } \B y\sim p_{y|x}, \B x=\B 0 }\cr
 & = \sum_{d=0}^n \min(1,2^{nR} g(d))  q_n(d;P\sub{bit})
}
\e{OOT6}
$$
where we used \e{ADO2bb}. Then the result is obtained by exploiting the equivalences $g(d)=P\sub{FA}(d)$, and $P\sub{MD}(d)=\sum_{i=d+1}^nq_n(i;P\sub{bit})$.
\end{IEEEproof}

\begin{IEEEproof}[Proof of Theorem~\ref{theo801}]
Note that the results of Corollary~\ref{hdTEO5b} provide simple approximations to estimate the value $d_0$ and the contribution $P\sub{MD}(d_0)$. We then want to identify a $O(n^{-1})$ approximation to the summation in \e{OOT2}. To this aim we preliminarily rewrite the bound in the form $\overline{P}_e=\alpha+\beta$ where $\beta=P\sub{MD}(d_0)$ and 
$$
\alpha = \sum _{d=0}^{d_0}  e^{n[u_n(d)-v_n(d)+R\ln(2)]} 
\e{WA2}
$$
with $v_n(d) = -\fract1n\ln P\sub{FA}(d)$, for which a $O(n^{-2})$ approximation is available from \e{HG3b}, and with
$$
\eqalign{
u_n(d) & = \fract1n\ln q_n(d;P\sub{bit}) \cr
 & = \fract1n \ln \left(\frac{\Gamma(n+1)}{\Gamma(n\!-\!d\!+\!1)\Gamma(d\!+\!1)}\right) + \ln(1\!-\!P\sub{bit}) - \frac{d}n \delta_0\cr
& = h(\lambda)\ln(2)+ \ln(1-P\sub{bit})-\lambda\delta_0 - \frac{\ln(2\pi n)}{2n} \cr
 & \qquad - \frac{\ln(\lambda(1-\lambda))}{2n} + O(n^{-2})
}
\e{WA4}
$$
where $\lambda=\frac dn$, and whose approximation was derived from the asymptotic expression \cite[eq.~6.1.41]{AbramowitzStegun68}. Use of the above $O(n^{-2})$ approximations provide the desired $O(n^{-1})$ approximation for $\alpha$, but they do not solve the closed form evaluation of the summation in \e{WA2}. To this end, we observe that, for a fixed value of $n$, the functions $v_n(d)$ and $u_n(d)$ can be interpreted as functions of $\lambda=\frac dn$. Their derivative with respect to $\lambda$ can also be identified (at least approximately) by exploiting Corollary~\ref{hdTEO5b} and \e{WA4}, to have
$$
\eqalign{
v_n'(\lambda) & = -\ln\left(\frac{1-\lambda}{\lambda}\right)+ \frac1n\frac{1-3\lambda+\lambda^2}{\lambda(1-\lambda)(1-2\lambda)} + O(n^{-2})\cr
u_n'(\lambda)
 & =  \ln\left(\frac{1-\lambda}{\lambda}\right) -\delta_0 - \frac{1-2\lambda}{2n\lambda(1-\lambda)} + O(n^{-2})\;.
}
\e{WA8}
$$
The idea is then to approximate $\alpha$ by exploiting the (truncated) Taylor expansion of the function $g_n(\lambda)=u_n(\lambda)-v_n(\lambda)$, namely
$$
g_n(\lambda)  = g_n(\lambda^*) + g_n'(\lambda^*) (\lambda-\lambda^*) + O((\lambda-\lambda^*)^2)\;,
\e{WA9}
$$
for some $\lambda^*$. A sensible choice is to choose $\lambda^*$ as the solution to $v_n(\lambda^*)=R\ln(2)$. With some effort it can also be verified that the $O((\lambda-\lambda^*)^2)$ contribution at the exponent corresponds to an $O(n^{-1})$ contribution to $\ln(\alpha)$, which is ensured by the fact that the polylogarithm series $\sum_{d=0}^{d_0} e^{-g' d}d^2$ is $O(1)$ for $d_0=\lceil n\lambda^*\rceil=O(n)$. Therefore, by exploiting the truncated Taylor expansion \e{WA9} and the $O(n^{-2})$ approximations for function $g_n(\lambda)$ and its derivatives, we obtain
$$
\frac1n\ln(\alpha)  =  u_n(\lambda^*) + \frac1n\ln\left(\frac{1-e^{-(d_0+1)g}}{1-e^{-g}} \right)+O(n^{-2})
\e{WA12}
$$
where $g = g_n'(\lambda^*)$. Note that the contribution $e^{-(d_0+1)g}$ can be included in the $O(n^{-2})$ factor since $d_0$ is $O(n)$, and that a $O(n^{-1})$ approximation of $g$ can be used, which, from inspection of \e{WA8}, provides
$$
\frac1n\ln(1-e^{-g}) = \frac1n\ln\left(1-\frac{(1-P\sub{bit})(\lambda^*)^2}{P\sub{bit}(1-\lambda^*)^2}\right)+ O(n^{-2})\;.
\e{WA16}
$$
From the second of \e{HG3b} and by considering that $P\sub{MD}$ is a decreasing function of $\lambda$ we further obtain
$$
\frac1n\ln(\beta) \lesssim u_n(\lambda^*) + \frac1n\ln\left(\frac{(1-\lambda^*)P\sub{bit}}{\lambda^*-P\sub{bit}}\right)+ O(n^{-2})\;.
\e{WA20}
$$
Equation \e{WA30} is finally obtained by putting altogether the above results.
\end{IEEEproof}

\begin{IEEEproof}[Proof of Theorem~\ref{theo8c2}]
We are interested in the bound on rate, in which case \e{WA30} must be interpreted as a constraint on $P_e$. For $n=\infty$, the first of \e{WA32} reveals that $\lambda=P\sub{bit}$. Therefore, for limited $n$ we are interested in the Taylor expansion at $P\sub{bit}$. For the first two expressions in \e{WA32} we find
$$
\eqalign{
w_n^{(0)}(\lambda) & = - \frac{\ln(2\pi n)}{2n} -\frac{(\lambda-P\sub{bit})^2}{2W} + O((\lambda-P\sub{bit})^3)\cr
(w_n^{(1)}(\lambda))^2 & = \frac{(\lambda-P\sub{bit})^2}{W} + O((\lambda-P\sub{bit})^3)
}
\e{OI2}
$$ 
with $W=P\sub{bit}(1-P\sub{bit})$. By following the rationale of Theorem~\ref{theo8} leading from \e{LX2}-\e{LX4} to \e{LX6}, we obtain
$$
\lambda = P\sub{bit} +\sqrt{\frac Wn} Q^{-1}(P_e) + O(1/n)\;.
\e{OI4}
$$
The normal approximation is finally derived from $\underline{R}=w_n^{(2)}(\lambda)\log_2(e)$ by using the last of \e{WA32}, and specifically by using \e{OI4} in the $O(n^{-1})$ approximation
$$
w_n^{(2)}(\lambda) =  [1-h(\lambda)]\ln(2)  +\frac{\ln(2\pi n)}{2n} + O(1/n)\;.
\e{OI6}
$$
\end{IEEEproof}

\bibliographystyle{IEEEtran}
\bibliography{fb}

\begin{IEEEbiography}[
]{Tomaso Erseghe} was born in 1972.  He received a Laurea (M.Sc degree) and a Ph.D. in Telecommunication Engineering from the University of Padova, Italy in 1996 and 2002, respectively. From 1997 to 1999 he was with Snell~\&~Wilcox, an English broadcast manufacturer. Since 2003 he has been an Assistant Professor (Ricercatore) at the Department of Information Engineering, University of Padova. His current research interest is in the fields of coding bounds in the finite blocklength regime, distributed algorithms for telecommunications, and smart grids optimization. His research activity also covered the design of ultra-wideband transmission systems, properties and applications of the fractional Fourier transform, and spectral analysis of complex modulation formats.
\end{IEEEbiography}
\vfill

\end{document}